\documentclass[journal]{IEEEtran}

\usepackage[utf8]{inputenc}
\usepackage{color}
\usepackage{xcolor}
\usepackage{array}
\usepackage{verbatim}
\usepackage{float}
\usepackage{amsmath}
\usepackage{amsthm}
\usepackage{amssymb}
\usepackage{graphicx}
\usepackage{longtable}
\usepackage{multirow}
\usepackage{booktabs}
\usepackage[unicode=true,
bookmarks=false,
breaklinks=false,pdfborder={0 0 1},colorlinks=false]
{hyperref}
\hypersetup{
	colorlinks,bookmarksopen,bookmarksnumbered,citecolor=blue,urlcolor=blue}
\usepackage{cite}

\usepackage{lipsum}
\usepackage{mathtools}
\usepackage{cuted}
\providecommand{\tabularnewline}{\\}
\usepackage{algorithmic}
\usepackage{longtable}

\floatstyle{ruled}
\newfloat{algorithm}{tbp}{loa}
\providecommand{\algorithmname}{Algorithm}
\floatname{algorithm}{\protect\algorithmname}

\makeatletter
\let\oldforeign@language\foreign@language
\DeclareRobustCommand{\foreign@language}[1]{%
	\lowercase{\oldforeign@language{#1}}}

\let\oldforeign@language\foreign@language
\DeclareRobustCommand{\foreign@language}[1]{%
	\lowercase{\oldforeign@language{#1}}}

\ifCLASSINFOpdf
\else
\fi

\@ifundefined{showcaptionsetup}{}{%
	\PassOptionsToPackage{caption=false}{subfig}}
\usepackage{subfig}

\usepackage{balance}

\ifCLASSINFOpdf
\else
\fi

\ifCLASSINFOpdf
\else
\fi

\def\ps@IEEEtitlepagestyle{%
	\def\@oddhead{\parbox[t][\height][t]{\textwidth}{\centering \scriptsize
			Personal use of this material is permitted. Permission from the author(s) and/or copyright holder(s), must be obtained for all other uses. Please contact us and provide details if you believe this document breaches copyrights.\\
			\noindent\makebox[\linewidth]{}
		}\hfil\hbox{}}%
	\def\@evenhead{\scriptsize\thepage \hfil \leftmark\mbox{}}%
	\def\@oddfoot{\parbox[t][\height][l]{\textwidth}{
			\vspace{-20pt}{\rule{\textwidth}{0.4pt}}\\ \footnotesize{\bf{\footnotesize\textcolor{red}{H. A. Hashim, "Advances in UAV Avionics Systems Architecture, Classification and Integration: A Comprehensive Review and Future Perspectives," Results in Engineering, vol. 25, pp. 103786, 2025.}}} doi: \href{https://doi.org/10.1016/j.rineng.2024.103786}{10.1016/j.rineng.2024.103786}\\\\
			\noindent\makebox[\linewidth]
		}\hfil\hbox{}}%
	\def\@evenfoot{\MYfooter}}

\makeatother
\begin{document}
	\bstctlcite{IEEEexample:BSTcontrol}

\title{Advances in UAV Avionics Systems Architecture, Classification and Integration: A Comprehensive Review and Future Perspectives}

\author{Hashim A. Hashim\\
	Department of Mechanical and Aerospace Engineering, Carleton University,\\
	Ottawa, ON, K1S-5B6, Canada (e-mail: Hashim.Mohamed@carleton.ca)
	\thanks{This work was supported in part by the National Sciences and Engineering Research Council of Canada (NSERC), under the grants RGPIN-2022-04937.}
	\thanks{H. A. Hashim is with the Department of Mechanical and Aerospace Engineering, Carleton University, Ottawa, ON, K1S-5B6, Canada (e-mail: Hashim.Mohamed@carleton.ca).}
}



\maketitle
\begin{abstract}
Avionics systems of an Unmanned Aerial Vehicle (UAV) or drone are
the critical electronic components found onboard that regulate, navigate,
and control UAV travel while ensuring public safety. Contemporary
UAV avionics work together to facilitate success of UAV missions by
enabling stable communication, secure identification protocols, novel
energy solutions, multi-sensor accurate perception and autonomous
navigation, precise path planning, that guarantees collision avoidance,
reliable trajectory control, and efficient data transfer within the
UAV system. Moreover, special consideration must be given to electronic
warfare threats prevention, detection, and mitigation, and the regulatory
framework associated with UAV operations. This review presents the
role and taxonomy of each UAV avionics system while covering shortcomings
and benefits of available alternatives within each system. UAV communication
systems, antennas, and location communication tracking are surveyed.
Identification systems that respond to air-to-air or air-to-ground
interrogating signals are presented. UAV classical and more innovative
power sources are discussed. The rapid development of perception systems
improves UAV autonomous navigation and control capabilities. The paper
reviews common perception systems, navigation techniques, path planning
approaches, obstacle avoidance methods, and tracking control. Modern
electronic warfare uses advanced techniques and has to be counteracted
by equally advanced methods to keep the public safe. Consequently,
this work presents a detailed overview of common electronic warfare
threats and state-of-the-art countermeasures and defensive aids. Furthermore,
UAV safety occurrences are analyzed in the context of national regulatory
framework and the certification process. Lastly, databus communication
and standards for UAVs are reviewed as they enable efficient and fast
real-time data transfer.
\end{abstract}


\section{Introduction}\label{sec1}

\subsection{Motivation}
Unmanned Aerial Vehicles (UAVs) are commonly referred to as drones,
Uninhibited Aircraft Systems (UASs), or remotely piloted aircrafts,
and these terms will be used interchangeably. In broad terms, a UAV
is a flying vehicle with no human presence on board, which can be
either remotely piloted by a human operator or partially or fully
autonomous without remote human intervention. In recent years, there
has been a boom in UAV use in a variety of fields, including, but
not limited to, wildfire detection, civil infrastructure inspections,
precision agriculture, transportation, delivery, and Intelligence,
Security, and Reconnaissance (ISR) missions \cite{hashim2023uwb_ITS,yang2022hybrid,jonnalagadda2024segnet,wanner2024uav,ren2024k}.
Majority of the UAVs can be broadly classified into two categories:
fixed-wing and rotary-wing. Although fixed-wing UAVs allow for high
speed and heavy payloads, they are not suitable for stationary-like
missions that require prolonged hovering (e.g., filming industry and
building inspection). Rotary-wing UAVs (e.g., quadcopters), on the
other side, are perfectly fit for stationary-like missions, but they
are disadvantaged by low speed and smaller payloads. Consequently,
the most suitable type of UAV is application-dependent. The first
generation of UAVs, that date back to World War I (WW-I), were controlled
through simple inertial mechanical systems \cite{clark2000uninhabited}.
However, lack of complete aerodynamics understanding in addition to
the absence of the pilot to manually compensate for unmodelled dynamics
and sensor drifts led to poor performance. In the period between WW-I
and WW-II, the first radio-controlled aircraft were tested, and the
UAV aviation electronics (AVIONICS) packages were born \cite{clark2000uninhabited}.
\begin{figure*}
	\centering{}\includegraphics[scale=0.75]{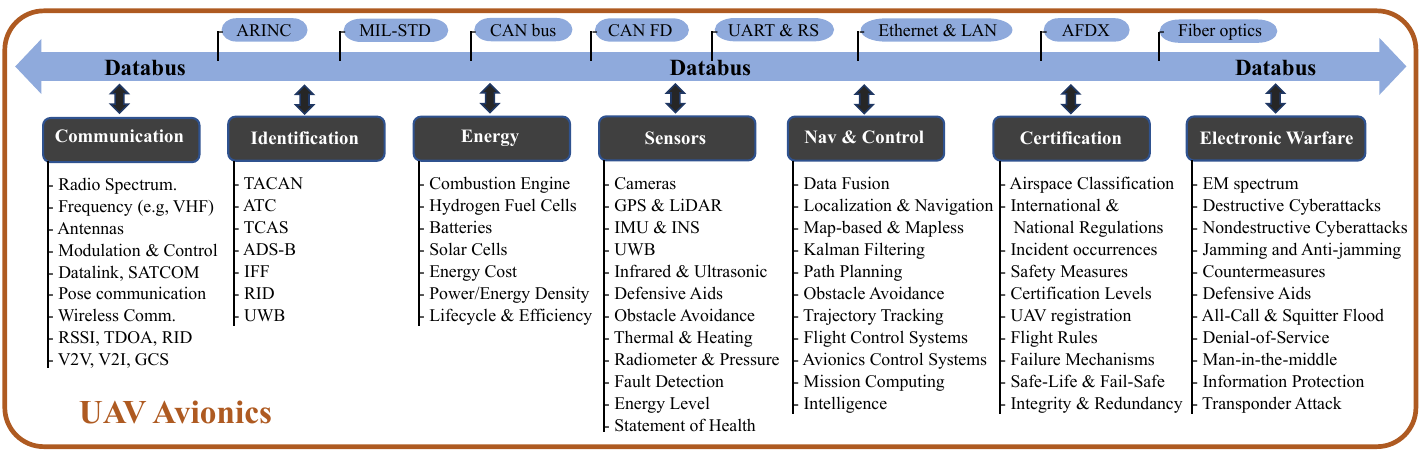}\caption{Illustrative diagram of UAV avionics elements.}
	\label{fig:UAV_Avionics}
\end{figure*}

Over the past decades, complex comprehensive avionics systems have
been developed and perfected through iterative improvement resulting
in modern cutting-edge UAVs able to execute sophisticated missions
for a multitude of industries and day-to-day tasks. UAV applications
are diverse and can require a single-agent UAV mission \cite{hashim2022ExpVTOL},
a homogeneous multi-agent UAV mission \cite{guo2019ultra}, or a mission
involving multiple heterogeneous agents (UAVs and ground vehicles)
\cite{cui2024resilient,hashim2022ExpVTOL}. While some UAV missions
are entirely dependent on a remote human operator others are either
partially or fully autonomous independent of remote human control.
Lack of onboard human operator necessitates precise and reliable UAV
avionics systems. Considering the fact that the number of UAVs carrying
out missions in the civil airspace is increasing exponentially, safe
navigation achieved through effective and standardized procedures
is paramount. Thus, it is crucial to ensure seamless and harmoniously
operation of all the UAV avionics systems that include the flight
control surfaces, UAV sensors, navigation and planning systems, communication
systems and power systems. Furthermore, safe and responsible UAV use
by general public, government bodies, and professional operators is
enforced by transportation agencies worldwide through rigorous certification
process and strict regulations. UAVs have powerful communication capabilities,
not only can they be connected to cellular networks, but they also
can enable terrestrial wireless communications by forming an assisted
communication system and acting as aerial base stations (BSs) or communication
access points \cite{soni2019performance}. Air-to-air communication
between UAVs and air-to-ground communication between UAVs and ground
stations are critical elements of UAV networks since they provide
the necessary means of identification and communication to achieve
the required tasks.

UAV's compact size and payload constraints make the energy resource
it carries in short supply. Consequently, the key component of mission
planning is optimization of the energy consumption and performance.
UAVs can use a variety of energy systems, such as batteries, fuel,
or renewable energy cells, and selecting the optimal energy system
and focusing of energy management allows to achieve extended UAV flight
duration and increased operational range \cite{yang2022hybrid}. Optimal
selection of the energy source and its efficient management allow
to minimize landing frequency for refueling or battery replenishment
enhancing UAV versatility. Successful completion of a scheduled mission
is highly dependent on the UAV's perceptual capabilities and the resulting
awareness of its current navigation states (starting point) in three-dimensional
(3D) space including location (position and orientation) in the six
degrees-of-freedom (6 DoF), speed, heading direction, and target destination.
Successful perception and navigation is built on four key pillars:
sensor selection, multi-sensor fusion, navigation techniques selection
(map-based and mapless), and robust estimator design. Finding a collision
free path in a cluttered environment requires careful path planning
from initial location to the final destination in 3D space while tackling
kinematic and dynamic constraints \cite{chen2021clustering,niu2019voronoi,prasad2020geometric}.
Locating the suitable obstacle free path in 3D space requires solving
the multi-objective Nondeterministic Polynomial-time (NP)-hard problem
that has no single optimal solution. Once the collision free path
is identified, trajectory control techniques are applied to track
the UAV along the desired route. Another critical component of UAV
avionics are the defense systems used to confront electronic warfare
threats, such as destructive and non-destructive cyberattacks, transponder
attacks and jamming threats, using state-of-the-art countermeasures
and defensive aids.

\paragraph*{Scope}The central objective of this review paper is to
present to the UAV research community with a comprehensive overview
of the UAV avionics systems architecture and classification including
key characteristics of each taxonomy. This work covers UAV communication
systems along with persistent challenges, as well as common warfare
threats and respective defense strategies. Another crucial UAV component
is its power source, hence this paper presents different energy sources,
energy-densities, power densities and their associated challenges.
Furthermore, typical identification systems used in UAVs and their
respective roles in communication systems, obstacle avoidance, and
electronic warfare are covered. The paper overviews perception sensors,
the role of sensor-fusion to building the navigation state, navigation
reliant on onboard sensors with and without external communication,
and navigation state estimation. Taxonomy of path planning techniques,
common methods applied for collision avoidance, and different control
approaches for trajectory techniques are presented. Finally, the UAV
regulations, safety protocols, airspace classification, and the certification
process necessary for an operator to carry out UAV missions are discussed.
Fig. \ref{fig:UAV_Avionics} provides a conceptual illustration of
UAV avionics components.

\paragraph*{Structure}The remainder of the paper is structured into
nine sections. Section \ref{sec:CommunicationSystems} presents communication
control systems, UAV antennas, communication of the UAV location (position
and orientation), and UAV-aided wireless communications. Section \ref{sec:IdentificationSystem}
discusses UAV identification systems. Section \ref{sec:EnergyStorageSystems}
summarizes different types of energy systems used by UAVs. Section
\ref{sec:NavigationSystems} discusses UAV perception and navigation
systems including sensor-fusion, map-based and mapless approaches,
and estimator design. Section \ref{sec:PlanningControl} presents
path planning, obstacle avoidance, and trajectory tracking. Section
\ref{sec:ElectronicWarfare} overviews UAV electronic warfare including
destructive and non-destructive cyberattacks, attacking transponders,
jamming and anti-jamming, and countermeasures and defensive aids.
Section \ref{sec:Databus} discusses databuses and their role in the
UAV avionics systems. Section \ref{sec:RegulationsCertification}
presents the regulations governing the use of UAVs and the associated
certifications. Finally, Section \ref{sec:Conclusion} summarizes
the work.

\begin{figure*}
	\centering{}\includegraphics[scale=0.43]{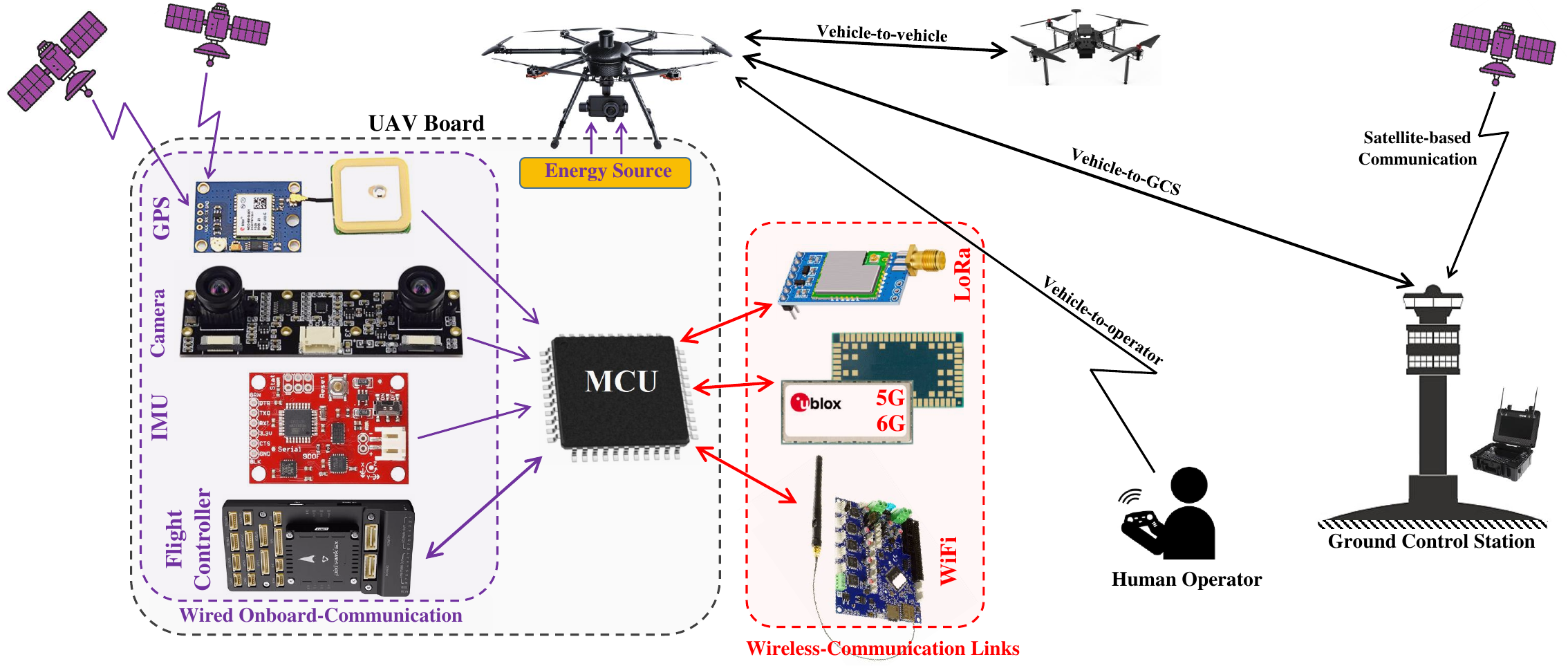}\caption{UAV communication systems: Wired, wireless, vehicle-to-vehicle, vehicle-to-infrastructure,
		and vehicle-to-operator. }
	\label{fig:UAV_comm}
\end{figure*}

\section{UAV Communication Systems\label{sec:CommunicationSystems}}

Practically, successful UAV missions rely on wireless communication
with the ground. UAV wireless communication is composed of sensor
communication (e.g., positioning), live image and video feeds, messages
to/from ground base station(s), connection with cell networks and
satellites, and collaboration with other air/ground vehicles where
communication links are established by means of commands. This section
reviews UAV communication systems, antennas, localization tracking,
and the use of UAV as aided-communication system. Fig. \ref{fig:UAV_comm}
presents illustrative diagram of UAV communication systems including
wired and wireless communication links, vehicle-to-vehicle, vehicle-to-infrastructure,
and vehicle-to-human operator. 

\subsection{Communication Control Systems}

Robust communication network for UAVs requires a variety of communication
modules and protocols. Communication control systems of UAVs are challenged
by multiple factors, such as high speed of UAV maneuvering, fuel/energy
consumption, air traffic density, UAV position and orientation, and
others. The essence of UAV external communication is the transmission
and reception of information in the form of data. Multiple antennas
distributed over the UAV body provide means of communication for military
and civilian applications. These antennas propagate information at
different frequencies of the electromagnetic spectrum complicating
the UAV communication \cite{hashim2023uwb_ITS,meng2023uav}. UAV antennas
communicate using the radio spectrum (30 Hz to 3,000 GHz), where the
typical ranges include Extremely Low Frequency (ELF) 30-300 Hz, Ultra
Low Frequency (ULF) 300-3000 Hz, Very Low Frequency (VLF) 3-30 kHz,
Low Frequency (LF) 30-300 kHz, Medium Frequency (MF) 0.3-3 MHz, High
Frequency (HF) 3-30 MHz, Very High Frequency (VHF) 30-300 MHz, Ultra
High Frequency (UHF) 0.3-3 GHz, Super High Frequency (SHF) 3-30 GHz,
Extra High Frequency (EHF) 30-300 GHz, and Tremendously High Frequency
(THF) 300--3,000 GHz. A significant amount of microwaves travel through
the sky, and therefore, to avoid interference, UAV communication is
enabled by allocating specific frequency ranges and modulation patterns
to each UAV antenna. Modulation, the process of converting low-frequency
periodic signals to/from high-frequency periodic signal, is commonly
employed to (i) reduce signal wavelength and in turn the required
size of an antenna, and (ii) minimize signal interference. Two main
types of modulation are utilized in UAV communications: analog (continuous)
and digital \cite{tacstan2021performance,li2019wavelet}. Analog modulation
is exemplified by Amplitude Modulation (AM), Phase Modulation (PM),
and Frequency Modulation (FM). Popular digital modulation approaches
include Amplitude Shift Keying (ASK), Phase Shift Keying (PSK), and
Frequency Shift Keying (FSK). UASs use Control and Non-Payload Communication
(CNPC) links to guarantee safety of their missions \cite{darsena2018equalization}.
CNPC links support secure two-way communications, low-latency, low
data rate, and reliable information propagation necessary for the
exchange of safety-critical data between UASs, manned aircrafts, and
Ground Control Stations (GCSs). The data transmitted using CNPC links
typically include control commands from GCSs, air traffic control
near airports, UAV status report to GCSs, and object avoidance information.
Datalinks facilitate telecommunication between two or multiple nodes,
including, gateway nodes, mobile terminals, terrestrial BSs, wireless
sensors, and others. In the event of complete BS malfunction or BS
offloading, datalinks can enable direct mobile-UAV communication.
Furthermore, datalinks support UAV-UAV and UAV-gateway wireless backhaul
\cite{li2018placement,zeng2016wireless}. 

\subsection{UAV Antennas}

\paragraph*{Antennas Role and Tracking Systems}UAVs without antennas
would be flying blind, unable to take advantage of any information
except for its pre-coded instructions and data garnered by the onboard
sensors. Consequently, antennas are integral to the UAV platform,
and are responsible for transmitting and receiving signals to and
from the ground stations or other aircraft in the form of electromagnetic
waves. Modern UAVs utilize a wide variety of antennas, such as wire,
aperture, and microstrip antennas. UAV communication network is affected
by a variety of factors, including antenna design, resource management
platform, network architecture, software complexities, and others.
Successful UAV tracking from a GCS requires three components: (a)
azimuth angle (angle between the north vector and the UAV on the horizontal
plane); (b) elevation angle (angle between a horizontal line and the
UAV starting from the GCS); and (c) absolute range. In GPS-denied
regions, the azimuth can be defined via the compass bearing system
\cite{hashim2021ACC}. The tracking antenna system maximizes the radio
communication between a UAV and a GCS. In the telemetry communication
systems for UAVs, ground-to-air antenna tracking systems are more
popular for UAV tracking. These located on the ground systems adjust
their position to achieve optimal elevation and azimuth for seamless
communication and employ two different antennas, such as Yagi-Uda
(parasitic array antenna composed of two types of parasitic elements
acting as a reflector and a director), a bi-quad (modified dipole
antenna with the wire shaped as a diamond or a square), and/or a double
bi-quad antenna \cite{fusselman2023long}%
. Simplicity and power efficiency of the ground-to-air tracking antenna
systems explain their widespread use for UAV missions. Furthermore,
these systems do not necessitate knowledge of the air vehicle attitude
which is a significant benefit \cite{fusselman2021ultra}. Another
antenna type is a circularly polarized microstrip antenna array that
operates in the microwave frequency range and is commonly employed
to enhance the bandwidth for UAV ground-to-air transmission \cite{kelechi2021recent}.
Microstrips, planar in design, represent advancement in antenna technology
and are heavily used \cite{liang2019reconfigurable}. They are perfectly
fit for aircraft and UAVs due to their planar design, allowing for
microstrip incorporation into the aircraft surface \cite{zhang2020pattern}.

\paragraph*{Air-to-Ground Communication and Modulation}Air-to-ground
communication systems are rarely used for UAV missions. UAVs typically
utilize directional antennas to transmit signals which help to focus
and amplify signals over long distances, prevent interference, and
reduces the burden for detecting the Angle of Arrival (AoA) of the
arrived signal \cite{guo2020optimal}%
. In air-to-ground communication, the tracking antenna positioned
on-board of aircraft persistently maintains track of the GCS. Air-to-ground
communication can use single carrier modulation, such as Quadrature
Phase Shift Keying (QPSK), Amplitude and Phase-shift Keying (APSK),
Minimum Shift Keying (MSK), Orthogonal Frequency Division Multiplexing
(OFDM), M-ary Quadrature Amplitude Modulation (MQAM), Gaussian Minimum
Shift Keying (GMSK), Shaped-offset Quadrature Phase-shift Keying (SOQPSK),
and Single-Carrier Frequency Domain Equalization (SC-FDE) \cite{khawaja2019survey}.
Furthermore, air-to-ground communication can use multi carrier modulation,
such as OFDM. All of the above-discussed antennas can enable UAV tracking
when used in conjunction with several techniques, such as monopulse,
Global Positioning System (GPS), Received Signal Strength (RSS) Indicator
(RSSI), and Time Difference of Arrival (TDOA) \cite{khawaja2019survey}.

\subsection{Position and Orientation Communication Tracking}

Safety of the airspace is highly dependent on the ability to accurately
track position and orientation of aircraft. In this subsection we
will discuss RSSI, TDOA, GPS and monopulse methods utilized together
with the data communicated via antennas to localize the aircraft of
interest. Firstly, let us look at RSSI - a measure of how well a device
hears a signal. Consequently, RSSI can be employed to calculate the
distance between the UAV and the tracking antenna using signal strength.
Moreover, RSSI allows to determine UAV flight direction, namely, a
UAV antenna transmits signal to the base stations and the direction
is established based on the progression of signal strength at the
receiver nodes \cite{khan2021rssi}. Use of RSSI in conjunction with
tracking antennas is characterized by such advantages as low cost,
energy consumption, and computational and memory requirements. However,
the main disadvantage is potentially unreliable position estimation.
\cite{spyridis2021modelling}. TDOA is another tracking method that
estimates transmitter location based on the differences in arrival
times by correlating signals received at different ground modes \cite{ouyang2020cooperative,hashim2023uwb_ITS}.
The TDOA approach calculates the time difference in wave arrival at
various sensor locations by utilizing signal and clock timing, along
with wave comparisons \cite{hashim2023uwb_ITS,hashim2023uwbIROS}.
In this approach, the UAV is typically equipped with a tag antenna,
whose position is unknown and requires estimation, while the tag communicates
with anchor antennas, which are usually fixed with known positions.
TDOA UAV tracking benefits from: simplicity in implementation, low
cost, and compact sensor size. Nonetheless, TDOA performance could
degrade if the signal bandwidth reduces \cite{xu2020three}. Moreover,
TDOA has another limitation which is the necessity for highly accurate
time synchronization. This is a serious limitation due to the trade-off
between accurate time synchronization and signal bandwidth \cite{ouyang2020cooperative}.

UAVs can also be tracked using a GPS module, very popular among ground
vehicles \cite{hashim2021ACC}. GPS communication between satellites
and an onboard antenna provide the UAV with its position and linear
velocity with respect to a reference coordinate while the orientation
can be extracted using the onboard Inertial Measurement Unit (IMU).
Note that various approaches (e.g., haversine distance and bearing
formula), several parameters can be calculated such as spherical trigonometry,
azimuth and elevation angles, and the absolute distance of the UAV
from the tracking station \cite{shah2017detecting}. Common challenges
associated with the use of GPS for UAV tracking are multipath ranging
errors, interference, and signal loss (e.g., indoors) \cite{hashim2021ACC,hashim2021_COMP_ENG_PRAC}.
To ensure success of every UAV mission, on-board computers are typically
loaded with backup algorithms reliant on sensor fusion (e.g., IMU
and camera) to estimate UAV navigation until strong GPS signal is
restored. The final tracking method discussed in this subsection is
monopulse, widely deployed in radar technology. Monopulse can be used
to find angular displacement of a UAV via array antennas and beam
forming \cite{song2021robust}. These antennas create a beam and the
signal reflected from the UAV becomes stronger as the UAV approaches
center of the beam and weaker as the UAV moves farther away from the
center of the beam.

\subsection{UAV-aided Wireless Communications}

High-speed wireless communication systems are essential for efficient
use of UAVs. UAVs can be employed as a \textit{UAV-aided wireless
	communication system} to provide wireless connectivity to other UAVs
or devices located in zones without infrastructure coverage, for instance,
areas devastated by natural disasters or areas that experience shadowing
by urban or mountainous terrain \cite{merwaday2015uav,zeng2016wireless}.
This type of communication is known as UAV-based Low-altitude Platforms
(LAPs) often utilized as a backup solution for High-altitude Platforms
(HAPs). An HAP system consists of an aircraft operating in the stratosphere
and transmitting signals to the Earth over long distance. This means
of communication offers a wider coverage and longer endurance when
compared to UAV-based LAPs \cite{xu2021survey}. Nevertheless, UAV-based
LAPs are cost effective, can enhance the performance of short-distance
communication, and could allow for adaptive communication \cite{zeng2016wireless,akram2020uav}.
The use of UAV-aided wireless communications can be divided into three
distinct categories: (i) UAV-aided ubiquitous coverage \cite{soni2019performance},
(ii) UAV-aided relaying \cite{alnagar2021q}, and (iii) UAV-aided
information dissemination and data collection \cite{zeng2016wireless}.
UAV-aided ubiquitous coverage employs UAVs to ensure that an established
communication infrastructure provides seamless wireless coverage.
To illustrate, a UAV could act as a backup station for a malfunctioning
or overloaded Base Station (BS). In UAV-aided relaying, UAVs are used
to enable wireless communication between distant users or user groups
that lack a direct communication link (e.g., military applications).
In UAV-aided information dissemination and data collection, one or
more UAVs are deployed to collect or transmit information through
a network of wireless devices (e.g., precision agriculture).

\section{Identification Systems\label{sec:IdentificationSystem}}

\paragraph*{Common Identification Systems}An identification system
is critical for the safe operation of both manned and unmanned vehicles
serving in civilian and military applications. An identification system
consists of a transponder, a radio frequency device that transmits
a signal in response to receiving an interrogating signal, and ensures
safe navigation and collision avoidance. More specifically, transponders
in aircraft are responsible for three tasks: localization (absolute
distance and azimuth angle), separation, and identification of aerial
vehicles \cite{Nato2012,sciancalepore2019sos}. Radar-like techniques
are typically employed to locate an aircraft. Typically, once a manned
the aircraft is in the radar’s range for interrogation, identification
is accomplished by assigning a special code to the aircraft via Air
Traffic Control (ATC) which is afterwards transmitted to the ground
station in addition to any other necessary information \cite{donmez2020handling}.
In this case, directional antennas are employed to broadcast interrogation
signals and receive responses. Time of of signal transmission and
return enables aircraft range measurement while the direction in which
the antenna is pointing when the signal is received is used to determine
the aircraft azimuth angle. For the purposes of separation and identification
of aerial vehicles, communication-like techniques are utilized. The
aircraft communication responses include identifying information,
including aircraft altitude \cite{donmez2020handling}. According
to the NATO definition, the transponder is a transmitter/receiver
device that is designed to transmit a response signal when legitimately
interrogated \cite{Nato2012}. Typical aircraft identification system
employs a number of different interrogators and transponders that
work together to guarantee safe mission completion: Distance Measurement
Equipment (DME) with UHF range, Tactical Air Navigation (TACAN) with
UHF range, ATC with LF range, Traffic Collision Avoidance System (TCAS)
with UHF range, Automatic Dependent Surveillance-broadcast (ADS-B)
with UHF range, and Identification Friend or Foe (IFF) with UHF range
\cite{moir2013civil}.

\paragraph*{Function of Identification Systems}DME provides the user
aircraft with distance from the station measurement. TACAN supplies
the aircraft with bearing (azimuth angle) and distance to the ground
or a vessel navigation through the water. ATC assists in aircraft
identification on air traffic control radars. TCAS reduces the incidence
of Mid-air Collisions (MACo) between aircrafts, monitors the airspace
and aircraft equipped employing a transponder, and warns the pilot
of other aircraft navigating in the vicinity and presenting a risk.
ADS-B helps with tracking aerial vehicles by defining the vehicle
position using satellite navigation or other positioning sensors.
IFF is one of the earliest identification systems involving a transponder
that was developed by the Germans during WW-II around 1941 to distinguish
friendly and hostile aircraft and the device was called German FuG
25A Erstling \cite{scott2015aviation}. Since then, transponder research
and technology for aerial vehicles have expanded to produce a variety
of transponders designed for a variety of applications, such as DMEs,
ADS-B, TACAN, and TCAS. IFF employs a transponder to listen for an
interrogation signal and then sends a signal which identifies the
broadcaster. Nowadays, IFF is essential for both civil and military
identification and is compatible with ATC. The ADS-B allows to broadcast
messages containing UAV location obtained by the GNSS along with other
information collected by the on-board measurement units to ATC and
other aerial vehicles. ADS-B has two operation modes, ADS-B In and
ADS-B Out. The ADS-B Out mode periodically reports UAV position, velocity,
and altitude to the ATC and the nearby aircraft. The transmitted data
acts as an equivalent of a radar display and no external action is
required. The ADS-B In mode allows for information exchange with the
other vehicles and ATC related to Flight Information Service - Broadcast
(FIS-B) \cite{sciancalepore2019sos}.

\begin{figure}
	\begin{centering}
		\includegraphics[scale=0.37]{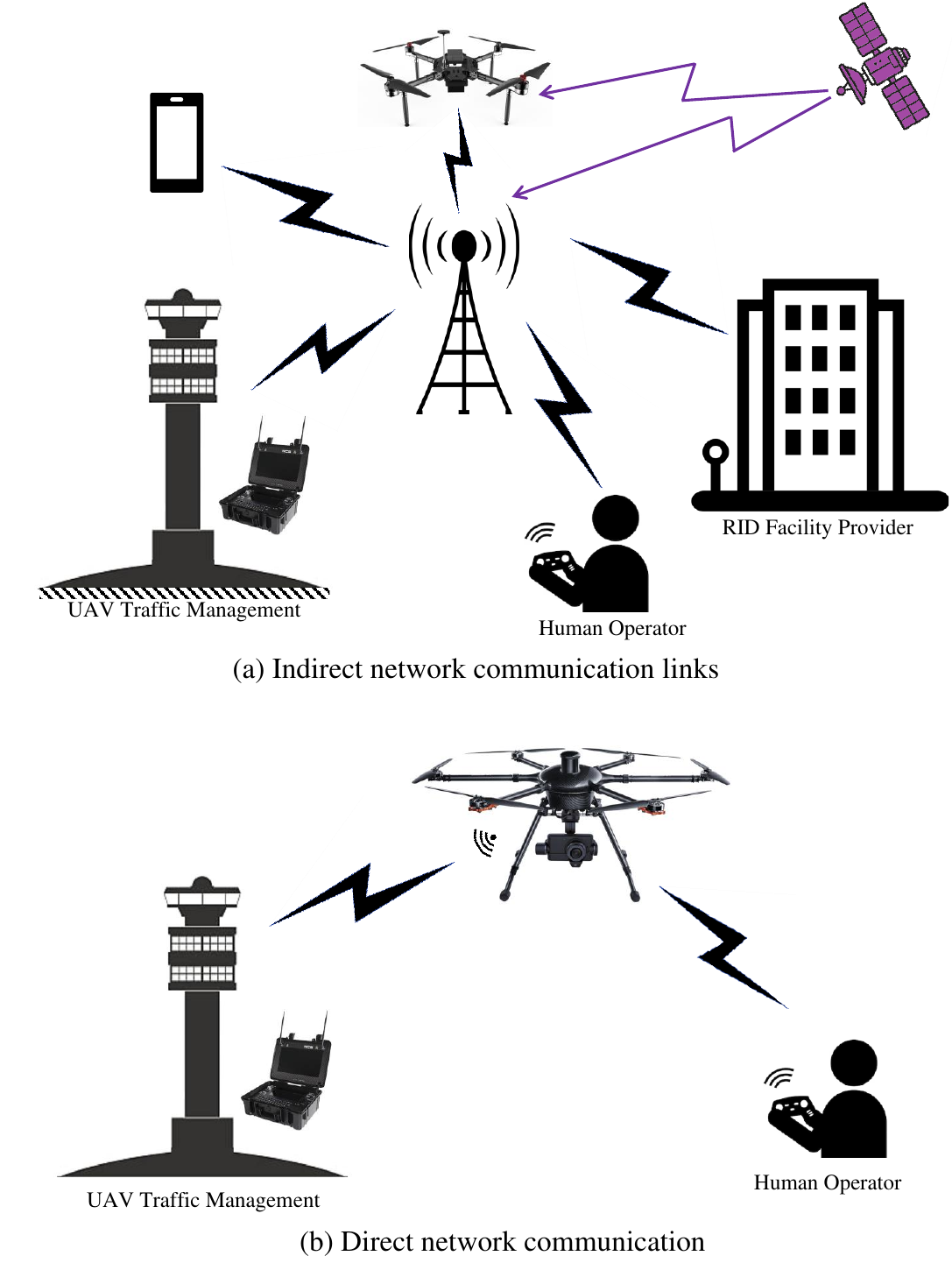}
		\par\end{centering}
	\caption{\label{fig:RID-Technologies}RID Communication Technologies.}
\end{figure}

\paragraph*{RID Systems and Network Broadcasting Technologies}There
exist many methods of UAV identification and the country of operation
determines the approach used. Radio frequency is a growing technology
that UAVs can utilize to broadcast their Remote Identification (RID)
\cite{lv2021drone}. In the USA, RID modules consisting of a transceiver
are utilized to broadcast location, speed, altitude, and other important
information to the ground stations. Furthermore, it is important to
note that since March 2024 UAVs weighing 250 grams or more and all
UAVs used for commercial purposes must be compliant with RID requirements.
RID can be either built-in or can come a separate module, either way
it should be able to broadcast UAV ID, emergency status, velocity,
UAV location and altitude, control station location and elevation,
and a time stamp. RID not only ensures safe sharing of skies with
other aircraft, but also allows to achieve UAV swarm formations (see
Fig. \ref{fig:RID-Technologies}). As part of RID, radio-frequency
transmitters can be used to broadcast periodical messages to advertise
UAV presence for other devices in the network. Network broadcasting
is a popular data transmission technique of the internet and services.
Devices can access UAV ID through distributed data where the most
popular network broadcasting technologies are cellular and satellite
networks (SATCOM) \cite{whitley2020unmanned} approximately UHF range.
UAV RID is still to be implemented with SATCOM technology. The currently
employed network broadcasting model requires the drone to be subscribed
in the network of a mobile service or mobile services, normally cellular
network \cite{shoufan2021esim}. The user must provide information
about the UAV by registering with an embedded Subscriber Identification
Module (eSIM) service by transmitting a request to the Mobile Network
Operator (MNO). Next, the MNO verifies the provided information and
communicates it to the Government Telecommunication Authority (GTA),
requesting activation code and PIN to finalize the UAV registration
\cite{belwafi2022unmanned}. In China and Russia, ADS-B system takes
care of this function \cite{wanner2024uav,michaelides2023challenges}.
Identification of UAVs is critically important since it allows to
manage the controlled airspace by preventing unauthorized entrances
which in turn ensures safety. UAVs can employ TCAS and ADS-B to realize
the collision avoidance system \cite{chand2017sense} necessary for
UAV swarm formations.

\paragraph*{Identification and Positioning with UWB Radio Technology}Ultra-wideband
(UWB) radio with SHF range is another method of UAV identification,
which is increasingly researched and used for UAV positioning in 3D
space \cite{hashim2023uwb_ITS,dwek2019improving,guo2019ultra,hashim2023uwbIROS}.
UWB transceiver technology, in addition to its low-cost implementation,
has multiple advantages making it an optimal fit for a variety of
applications (e.g., smartphones and smart watches) \cite{hashim2023uwb_ITS,liu2023real,ramirez2023joint}.
UWB signal is characterized with large bandwidth and short life-time
resulting in good positioning accuracy (10 centimeters in a range
of 100 meters) \cite{hashim2023uwb_ITS}. Moreover, UWB signal is
distinguished by power efficiency, fast communication speed, short
wavelength, robustness against multipath interference which enables
UAV positioning in GPS-denied regions, and ability to penetrate obstacles
facilitating Line-of-Sight (LOS) and Non-Line-of-Sight (NLOS) communication
\cite{hashim2023uwb_ITS,hashim2023uwbIROS}. Using UWB technology
for UAV identification requires two main elements: anchors and tags
\cite{hashim2023uwb_ITS}. UWB anchors constitute fixed (e.g., attached
to structures or walls) or mobile (e.g., attached to smartphones or
UAVs that have access to GPS) sensors whose position is known \cite{hashim2023uwb_ITS}.
Tags, on the other hand, represent moving sensors whose position is
unknown, such as a UAV equipped with a UWB unit. Each UWB-tag communicates
with multiple UWB-anchors to establish the tags position in 3D space
by transmitting messages containing UAV identification and a timestamp
recorded using the clock of the transceiver that is transmitting the
signal \cite{hashim2023uwb_ITS}.

\section{Energy Storage Systems\label{sec:EnergyStorageSystems}}

Energy generation and management in UAVS is not only a technical consideration,
but rather a fundamental element that shapes the operational viability
and sustainability. Hence, Energy Storage Systems (ESSs) are one of
the central components of UAV avionics. The market offers multiple
different power sources and ESSs for UAVs, such as petrol and diesel
combustion engines, Fuel Cells (FCs), batteries, solar cells, and/or
battery hybrid systems. ESSs and the power sources that comprise them
are evaluated and compared with respect to their energy and power
densities.

\subsection{Combustion Engine}

Powerful unmanned aerial transportation systems characterized with
long duration flight are in great demand and petrol-powered UAVs have
been developed to fill this niche \cite{FlyAbility2022,abdulsada2022design,koley2024electrochemistry}.
However, it should be noted that gasoline-powered UAVs has not been
developing at the same pace as electric-powered UAVs which have seen
significant growth in both research and application. Petrol and diesel
engines both belong to the family of internal combustion engines.
An internal combustion engine is a well-established system composed
of a combustion chamber, an intake valve, fuel injectors, pistons,
and an exhaust valve. Note that petrol engines requires spark plugs,
while diesel fuel self-ignites under extreme pressure and does not
require spark plugs \cite{zangana2023investigated}. Furthermore,
petrol engines have lower efficiency than diesel engines \cite{zangana2023investigated}.
Meanwhile, diesel engines could be challenging to start in cold temperatures
creating a need for heating up the diesel before entering the chamber
\cite{selvan2022utilization}. Petrol-powered vehicles utilize Ethanol,
Methanol, Kerosene, or Liquefied Petroleum Gas (LPG) Propane, which
allow the UAV to fly for more than 24 hours \cite{yang2022hybrid}.
The main advantages of combustion engines are the gradual reduction
of UAV weight during flight due to fuel consumption, high energy density,
and longer flight time when compared to electric-powered UAVs. On
the other hand, combustion engines are large in size rendering them
unsuitable for smaller UAVs, they are also environmentally unfriendly,
and require continuous maintenance.

\subsection{Hydrogen Fuel Cells}

A FC is a device that converts chemical energy into electrical energy.
Notably, a FC ESS is capable of producing energy for a longer period
of time than batteries. FCs that use chemical energy of hydrogen can
be employed by UAVs and come in a variety of forms, for instance,
an Alkaline FC (AFC), a Phosphoric Acid FCs (PAFC), a Proton Exchange
Membrane (PEM) FC (PEMFC), a Solid Acid FC (SAFC), a High Temperature
FC (HTFC), and an Electric Storage FC (ESFC) \cite{taylor2022hydrogen}.
The AFCs are distinguished by simple structure and high energy efficiency.
However, AFCs have short operating life-cycle because of the eroding
and have environmental impacts \cite{hazri2021critical}. The principal
of operation of the PEMFC is similar to that of a battery system.
More specifically, the PEMFC consists of two electrodes,:an anode
and a cathode, separated by an electrolyte membrane. This way electron
flow is generated by delivering fuel to the anode and supplying oxidant
to the cathode leading to an electrochemical reaction. One of the
key benefits of hydrogen FCs, is the fact that they use hydrogen as
fuel and air as the oxidant, resulting in only two process byproducts:
water and air. Using PAFCs, the fuel is hydrogen while the electrolyte
is liquid phosphoric acid \cite{oh2023energetic}. PEMFCs could maintain
high efficiency of both energy and power densities provided that they
operate at low temperatures \cite{fan2023recent}. Consequently, PEMFCs
are the preferred choice for electric vehicle applications due to
their compact size, light weight, and rich power source, making them
a potential fit for UAVs. Overall, FCs are marked by a significantly
higher energy density than batteries. Curiously, in a recent study
a fixed-wing UAV using a FC was able to complete a flight of 24 hours
\cite{taylor2022hydrogen}.

To summarize, the main advantages of using FCs as an ESS for UAVs
are the high energy and power densities, absence of noise, no direct
pollution, and fast refueling. Furthermore, similarly to combustion
engines, the weight of the UAV reduces gradually during the flight
allowing for longer flight time. Meanwhile, the persistent challenges
of FC use for UAVs are their large size and heavy weight.

\begin{table*}[t]
	\begin{centering}
		\caption{\label{tab:BatteryCompare} Comparison between different type of batteries
			in terms of nominal cell voltage, life-cycle, cost, power and energy
			density, and charge/discharge efficiency.}
		\par\end{centering}
	\centering{}%
	\begin{tabular}{lcccccccccc}
		\toprule 
		Type & Alkaline & Li-Ion & Li-Po & Li-air & Li-S & Li-SoCl2 & Pb-acid & NiCad & NiMH & Zn-O2\tabularnewline
		\midrule
		\midrule 
		Nominal cell voltage (V) & 1.3-1.5 & 3.6-3.8 & 2.7-3 & 2.91 & 2.5-2.6 & 3.5 & 2.1 & 1.2 & 1.2 & 1.45-1.65\tabularnewline
		\midrule 
		Life-cycle & 500 & 400-1,200 & 500 & 700 & N/A & N/A & 350 & 2000 & 180-2,000 & 100\tabularnewline
		\midrule 
		Cost (\$US/Wh) & 2.2-2.4 & 0.9-1 & 2.2-2.4 & N/A & 0.6-0.9 & 0.5-0.6 & 0.6-0.7 & 2.5-3 & 0.8-0.9 & 0.3-0.4\tabularnewline
		\midrule 
		Charge/Discharge efficiency (\%) & 90 & 80-90 & 90 & 93 & N/A & 6-94 & 50--95 & 70-90 & 66-92 & 60-70\tabularnewline
		\midrule 
		Power density (W/kg) & 50 & 250-340 & 245-430 & 11,400 & 2,600 & 18 & 180 & 150 & 250-1,000 & 100\tabularnewline
		\midrule 
		Energy density (Wh/kg) & 89-190 & 100-265 & 100-265 & 11,000 & 2,510 & 500-700 & 30-40 & 40-60 & 60-120 & 442\tabularnewline
		\bottomrule
	\end{tabular}
\end{table*}

\subsection{Batteries}

Analogously to fuel cells, batteries are devices that convert chemical
energy to electrical energy. However, while fuel cells use fuel to
generate energy, batteries store the energy they require. Different
types of batteries are utilized for UAVs, for instance, Alkaline,
Lithium Ion (Li-Ion), Lithium Polymer (Li-Po), Lithium-air (Li-air),
Lithium-sulfur (Li-S), Lithium-Thionyl-chloride (Li-SoCl2), Lead acid
(Pb-acid), Nickel cadmium (NiCad), Nickel Metal Hydride (NiMH), and
Zinc Oxide (Zn-O2) \cite{ci2016reconfigurable}. One of the most popular
choices for UAVs are Li-Ion and Li-Po batteries due to their low-cost
value per unit \cite{zhang2022review}. Moreover, Li-Ion batteries
are compact, light-weight, and capable of supplying reasonable amount
of energy and power per unit of battery mass when compared with other
rechargeable batteries \cite{vidal2019xev}. Given same weight and
volume, Li-air batteries could provide higher energy density ($\sim$
5-10 times) than the Li-Ion batteries \cite{kim2012power,townsend2020comprehensive}.
However, the significant shortcomings of the Li-air batteries are
the limited number of discharge cycles, the slow rate of recharge,
and the high risk of damage in presence of water vapor. Li-S and Li-SoCl2
batteries are known for their higher energy density per kilogram (kg)
\cite{zhou2022formulating}. However, the Li-SoCl2 type is more expensive
than the Li-Ion and Li-Po. It is worth noting that Li-S batteries
are more cost-effective and are expected to receive wider application
in UAV ESSs in the near future. Battery suitability can be evaluated
and compared based on a multitude of factors, for example, weight,
volume, energy density, power density, cost per unit, life-cycle,
safety and maintenance, available power management techniques (state
of health and charge), and others. Each of the above-listed characteristics
plays a critical role for selecting the best option for optimal UAV
operation. Energy density defines the maximum UAV range, while power
density specifies the UAV acceleration capabilities. Life-cycle measures
the number of times that the battery can be used before replacement.
Weight and volume affect the UAV flight range. Cost per unit plays
a critical role of battery production, availability, and affordability
(will it be affordable for the public? will it be cost-effective and
of interest for certain industries?). Table \ref{tab:BatteryCompare}
presents the main features of different types of batteries. The compared
characteristics include nominal cell voltage (V), life-cycle, unit
cost in terms of US dollars to watt hour (\$US/Wh), efficiency, power
density watt per kg (W/kg), and energy density (Wh/kg). The information
used for the comparison was collected from \cite{rajashekara2013present,kim2012power,townsend2020comprehensive,zhu2019far}.

Use of batteries for UAVs is very common due to multiple advantages
of this ESS, namely absence of noise and direct pollution, ease of
replacement of faulty parts due to multiple cell structure, and ease
of transport and recharge. Nonetheless, disadvantages are also present,
namely, limited number of recharge cycles and low energy density when
compared to FCs and petrol sources.

\subsection{Solar Cells and Solar Power}

Power as an ESS for aircraft and UAVs in particular is still in its
infancy, solar energy has significant potential for offering carbon-free
solutions for aerospace. There are three main technologies employed
for converting energy of the sun into electricity, namely, photovoltaic
(PV) cell, solar thermal collectors for heating and cooling (SHC),
and concentrated solar power (CSP). Thus far, PV cells are the only
technology utilized by aerospace applications. PV cells are devices
that directly convert sun rays into electrical energy which is used
for propulsion, powering of on-board systems, and the excess is used
to recharge batteries employed in the absence of or low sunlight conditions.
Particles of solar energy known as photos are absorbed by the semiconductor
material composing the PV cell \cite{dhaouadi2022modelling}. The
structure of a PV cell is made up of a positive and a negative layer,
forming a junction between them known as the p-n junction. Once a
sufficient amount of photons is absorbed by the semiconductor material
of the PV cell, electrons are forced to flow in one direction creating
an electric field at the p-n junction resulting in a flow of electricity.
In this simple and direct manner sunlight is converted into electric
current, which can be stored in a battery. The first solar powered
aircraft carried out a flight in 1974 \cite{boucher1985sunrise} where
the fuel system was composed of 4096 silicon PV cells and the operating
efficiency was near $11\%$. The Airbus Zephyr, the most advanced
solar powered UAS, operates in the stratosphere flying at an altitude
close to 70,000 feet \cite{hasan2021conceptual}. The wingspan of
the Zephyr is 25 meters which allows it to carry a large number of
solar panels, in turn, enabling continuous flight for long periods
of time. Typically, solar cells are installed on UAV fixed wings \cite{girlevicius2022evaluation},
and in order to achieve energy maximization a large surface of solar
panels is needed. This requirement makes PV cells unfit for small
scale UAVs. Furthermore, the solar panels require sunlight to operate
which is challenging in the event of limited exposure or complete
absence of direct sunlight. To summarize, although solar cells have
no direct operating costs like fuel or battery power, they require
complicated support systems, have a relatively high upfront integration
cost, and add a significant weight load to the UAV system.

\section{Perception and Navigation Systems\label{sec:NavigationSystems}}

UAV navigation is a process of self-vehicle localization which helps
the UAV to plan on how to safely and quickly proceed from the current
location to the target destination (motion planing and control). Successful
navigation mission completion requires the UAV to have access to its
own location, heading angle, and navigation speed which are commonly
referred to as UAV motion state (composed of attitude commonly known
as orientation, position, and linear velocity) in 3D space. The earliest
navigation methods included piloting, celestial navigation, and dead-reckoning.
Piloting method of navigation utilizes visual natural and human-made
landmarks (e. g., towns, rivers, lighthouses, or buoys) to enable
the vehicle's position determination. Celestial navigation relies
on position determination using sun, moon, stars, and planets. Dead
reckoning, classified as an inertial navigation approach, calculates
vehicle's current position based on the last known position, and advancing
it considering known or estimated linear velocity (integrated using
position information) over elapsed time and course. Modern navigation
has come a long way, and UAVs commonly rely on GNSS and SATCOM for
routing (for determining their position). State-of-the-art UAV navigation
techniques include: satellite navigation, inertial navigation, and
vision-based inertial navigation. While some methods are used as main
navigation techniques, others are employed as backup solutions. Safe
completion of a UAV mission is paramount, as such, employing multiple
navigation techniques at the same time ensures that reliable navigation
data can be obtained in all traveling conditions.

\begin{figure*}
	\centering{}\includegraphics[scale=0.53]{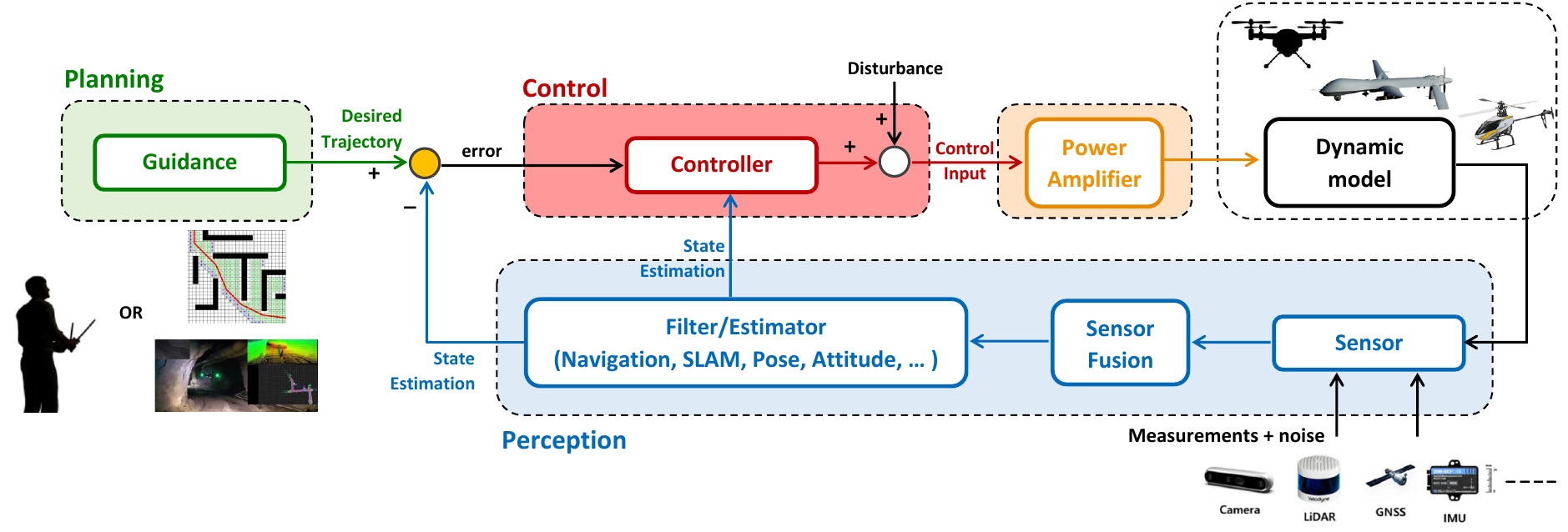}\caption{Conceptual diagram of perception, motion planning and control.}
	\label{fig:UAV_NavCont}
\end{figure*}

\subsection{Navigation Sensors and Multi-sensor Fusion}

A typical UAV is equipped with exteroceptive (for the external environment)
and proprioceptive (vehicle status and operation) sensors that enable
estimation of the UAV motion states. Commonly used sensors are GPS,
IMU (3-axis accelerometer, 3-axis gyroscope, and optional 3-axis magnetometer),
UWB tags and anchors, laser lightning and Light Detection and Ranging
(LiDAR), vision units (monocular, stereo, or RGB-D cameras), magnetic,
and ultrasonic sensors \cite{hashim2023uwb_ITS,hashim2021_COMP_ENG_PRAC,el2020inertial}.
GPS is a satellite-based sensor that supplies vehicle's position,
speed, heading/direction, and time information provided that signals
from at least four satellites are available \cite{hashim2023uwb_ITS,hashim2021_COMP_ENG_PRAC}.
Although GPS is widely used as a standard positioning system, GPS
is unavailable indoors and is subject to signal obstructions, multipath,
fading, and denial \cite{hashim2023uwb_ITS,hashim2021_COMP_ENG_PRAC}.
Furthermore, GPS does not provide vehicle's orientation. Other common
well-researched GPS shortcomings include satellite clock error, receiver
clock error, Ionospheric delay, Tropospheric delay, satellite orbital
(ephemeris) errors, receiver noise, and errors attributed to satellite
geometry. IMU is typically employed for the vehicle's orientation
determination. Although the use of IMU in conjunction with dead reckoning
can prove useful for short-path navigation, it is unreliable for long-path
navigation due to error drift and accumulation. UWB technology previously
discussed in Section \ref{sec:IdentificationSystem} has a concept
of operation similar to that of satellite-based positioning. UWB tag(s)
can enable the vehicle to localize its position in 3D space given
known fixed or moving anchors \cite{hashim2023uwb_ITS}. However,
UWB sensors are susceptible to measurement noise. Vision units determine
vehicle's position given earth-frame by tracking feature points using
two different instantaneous frames (stereo camera) or two consecutive
frames (monocular camera) \cite{hashim2021_COMP_ENG_PRAC}. It is
worth noting that vision units could face challenges in low-texture
environments and positioning accuracy degradation during high-altitude
flights. LiDAR is a distance measurement sensor, whose principal of
operation is analogous to that of radar but rather than using radio
waves LiDAR employs laser pulse (distance is calculated by recording
the time elapsed between emitting a pulse and receiving its reflection).
LiDAR is distinguished by high speed and accuracy. However, the high
accuracy is achieved by collecting large data volume that can cause
UAV on-board computer to overload, especially in unstructured areas
(e.g., machinery zones or shelving). Furthermore, the high cost of
the LiDAR unit makes it unfit for low-cost UAVs.

Combining information from multiple sensors allows to obtain comprehensive
knowledge of vehicle's motion state. It is achieved through multi-sensor
fusion, the process of data integration between different types of
sensors to enable state estimation of an augmented kinematical system
while improving the estimation accuracy. In navigation, multi-sensor
fusion is employed along with estimator or filter design to enable
full navigation determination. For instance, GPS cannot provide vehicle's
orientation information, while IMU cannot provide reliable information
of vehicle's position and linear velocity. Integrating GPS with IMU
(GPS-IMU fusion) results in an augmented kinematical system referred
to as navigation kinematics which allows for collecting reliable navigation
data. Examples of multi-sensor fusion include GPS-IMU \cite{girbes2021asynchronous},
vision-based navigation (vision unit + IMU) \cite{hashim2021_COMP_ENG_PRAC},
LiDAR-based IMU \cite{zhang2021lilo}, UWB-IMU \cite{hashim2023uwb_ITS},
and others. The estimation accuracy and reliability can be further
improved by combining multiple types of multi-sensor fusion. One of
the challenges of integrating data from multiple sources is the varying
frequency of data measurements. For instance, a typical IMU will supply
gyroscope and accelerometer measurements at a rate of 200 Hz, while
low-cost vision unit could supply images at a rate of 20 Hz \cite{hashim2021_COMP_ENG_PRAC,hashim2021ACC}.
Therefore, multi-sensor fusion data collection and processing algorithms
must account for frequency variations between different sensors.

\subsection{Map-based vs Mapless Navigation}

Map-based navigation involves determining location of a vehicle within
a predefined map or environment. A most common type of map-based navigation
is satellite-based navigation that involves fusion of onboard sensors,
namely, GPS and a 6-axis IMU (3-axis accelerometer and 3-axis gyroscope).
The fused data are forwarded to an estimator to identify the navigation
state of a vehicle, for instance, a UAV. The onboard GPS mounted at
the top of the UAV requires persistent access to the signal from at
least four different satellites \cite{hashim2023uwb_ITS}. GPS receiver
onboard a UAV listens to the radio signals broadcast by GPS satellites.
Distance from each satellite is accurately determined using precise
transmission time $t_{1}$ measured by satellite on-board atomic clocks
using the following formula:
\[
\text{Distance}=\text{velocity of light}\times\text{transit time}
\]
where transit time $=t_{2}-t_{1}$ with $t_{1}$ being time signal
at transmission and $t_{2}$ being the exact time of arrival. Note
that the GPS system is composed of three segments, namely space segment,
control segment, and user segment, where the U.S. Space Force develops,
maintains, and operates the space and control segments \cite{moir2013civil}.
Satellite-based navigation considers a predefined mapped areas. To
address indoor navigation missions and GPS-signal loss, Inertial-based
Navigation Systems (INS) can be adopted, such as UWB-IMU-based navigation,
LiDAR-based IMU, or vision-aided inertial navigation (vision unit
and a 6-axis IMU) \cite{hashim2021_COMP_ENG_PRAC,hashim2021ACC,hashim2023uwb_ITS,zhang2021lilo}.
The above-listed techniques consider unknown UAV localization problem.
In the event of a UAV traveling through an unknown environment mapless
navigation techniques are needed, for instance, Simultaneous Localization
and Mapping (SLAM) could be considered. SLAM is a challenging process
involving concurrent estimation of an unknown UAV state and and an
unknown map of the environment (navigation via mapless system) \cite{hashim2021T_SMCS_SLAM,Hashim2021AESCTE,hashim2020TITS_SLAM}.
If a vision-unit is involved in the estimation process, the localization
and mapping problem is denoted as Visual SLAM (VSLAM) \cite{Hashim2021AESCTE,hashim2020TITS_SLAM}.

\subsection{Estimator Design from Sensor Fusion}

\begin{figure}
	\begin{centering}
		\includegraphics[scale=0.69]{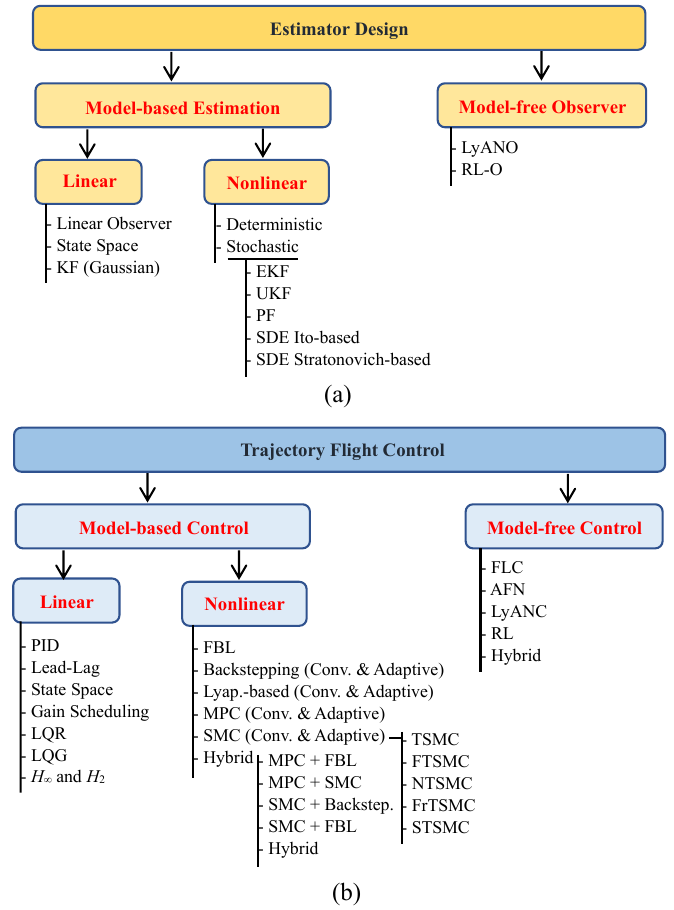}
		\par\end{centering}
	\caption{\label{fig:EstimatorControl}Classification of estimator and trajectory
		control design.}
\end{figure}

Unfortunately, full-state UAV navigation cannot be achieved by using
multi-sensor fusion measurements directly. Although UAV navigation
state can be reconstructed algebraically, it has been proved unreliable
due to measurement uncertainties \cite{hashim2021ACC}. Moreover,
navigation state is essential for controlling UAV motion and using
its direct algebraic reconstruction from multi-sensor fusion measurement
can lead to actuator failure and UAV destabilization due to sensor
reading drifts and noise \cite{hashim2021_COMP_ENG_PRAC,hashim2021ACC,hashim2023uwb_ITS}.
Consequently, an estimator design crucial for combining measurements
from multiple sensors, establishing error criteria, adaptively correcting
and smoothly estimating UAV navigation state, estimating UAV navigation
hidden-state (e.g., linear velocity), and robustly rejecting sensor
measurement uncertainties. Navigation estimators can be classified
as either deterministic or stochastic. Deterministic navigation estimators
are a good option when high-quality sensors are used, and examples
include state-space linear observer design or nonlinear complementary
filters \cite{hashim2020TITS_SLAM,hashim2021T_SMCS_SLAM}. Stochastic
navigation estimators can be linear and nonlinear. Linear-type stochastic
estimators include Kalman filters (KFs) \cite{feng2020kalman,odry2018kalman,odry2021open}.
Nonlinear-type stochastic navigation estimators include extended Kalman
filters (EKFs) \cite{potokar2021invariant} which require linearization
around nominal point, unscented Kalman filters (UKFs), Particle filters
(PFs) \cite{Khashayar2024Hashim,lou2023consider}, and Lyapunov-based
nonlinear complementary stochastic filters which use Stochastic Differential
Equations (SDEs) and adopt Ito's or Stratonovich's integrals to mitigate
noise stochasticity and address navigation nonlinearity \cite{hashim2021_COMP_ENG_PRAC,hashim2021ACC,hashim2023uwb_ITS}.
Ito's approach addresses white noise, while Stratonovich's approach
is applicable for colored noise \cite{hashim2018SO3Stochastic,hashim2019SO3Wiley}.
Lyapunov-based SDE Ito and Stratonovich-based filters are more computationally-efficient
and produce better results than KF, EKF, UKF, and PFs. The above-listed
estimators are model-based. Model-free based estimators include learning
based approaches (Lyapunov-based Adaptive Neural Observer (LyANO)
or Reinforcement Learning-based Observer (RL-O)) \cite{hashim2022NNCont}.
Fig. \ref{fig:EstimatorControl}.(a) illustrates estimator design
classification.

Fig. \ref{fig:UAV_NavCont} presents the process diagram of perception,
motion planning, and control of UAVs. On-board sensors (e.g., GPS,
IMU, LiDAR, camera) collect measurements corrupted with noise and
constant drifts. Next, sensor fusion algorithms combine different
sensor measurements and supply data to the filter (e.g., KF, EKF,
UKF, or nonlinear stochastic filter). The filter typically completes
three different tasks: filtering out measurement noises and drifts,
estimating UAV current location or navigation state, and observing
hidden states (e.g., UAV linear velocity in the 3-axis motion is typically
not available). At the next step, the guidance process completes motion
planning and generates the shortest obstacle-free trajectory. The
final link in the process is the controller that accesses the UAV
current state with respect to the desired state determined based on
the desired trajectory and thereafter applies the trajectory tracking
approach to guide the UAV to the desired destination.

\section{Path Planning, Collision Avoidance, and Control \label{sec:PlanningControl}}

\subsection{Path Planning}

Path planning and Collision Avoidance (CA) are the key elements of
UAV communication and trajectory control autonomy. Optimal selection
of a UAV path can reduce communication distance (see Section \ref{sec:CommunicationSystems}),
in turn enhancing the capacity performance (fly time and energy management).
Identifying the best UAV path is a complicated task due to the fact
that a large number of variables have to be optimized to produce the
most suitable trajectory form the current UAV pose to the desired
destination which is challenging making it NP hard problem. Also,
path planning is subject to a finite transition constraints. Additionally,
planning the optimal UAV path involves considering such crucial elements
as fuel status and limitations, nearby obstacles (potential collision
hazards and terrain type), obstacle avoidance strategies, path length,
optimality and robustness of the solution, cost and time-efficiency,
and system connectivity \cite{jones2023path}. These elements could
be time-variant and challenging to model. It is worth noting that
the planning process is further complicated by the fact that UAVs
navigate in 3D space as opposed to ground vehicles traveling in 2D.
Unfortunately, an exact algorithm for identifying the optimal path
exists for neither 2D nor 3D space navigation. Consequently, path
planning of the UAV environment is typically investigated through
one of the three strategies: cell decomposition, roadmap, and potential
field. Cell decomposition techniques represent the environment space
as a series of non-overlapping cells creating a navigable structure
for the UAV \cite{chen2021clustering}. Commonly applied cell decomposition
approaches include exact, approximate, and adaptive \cite{kyaw2020coverage}.
Exact cell decomposition approach structures the environment space
into non-overlapping polygon regions applying either trapezoidal or
boustrophedon decomposition \cite{pham2020complete}. This approach
has a limitation of not possessing uniform shape or structure of the
environment. Approximate cell decomposition, on the other hand, divides
the environment into a group of structured cells such that every cell
is described using a Cartesian coordinate form \cite{kyaw2020coverage}.
Finally, adaptive cell decomposition formulates the environment only
in zones where obstacles are present by applying decomposition recursively
once an obstacle is identified \cite{xiong2019path}. 

\begin{figure}
	\begin{centering}
		\includegraphics[scale=0.67]{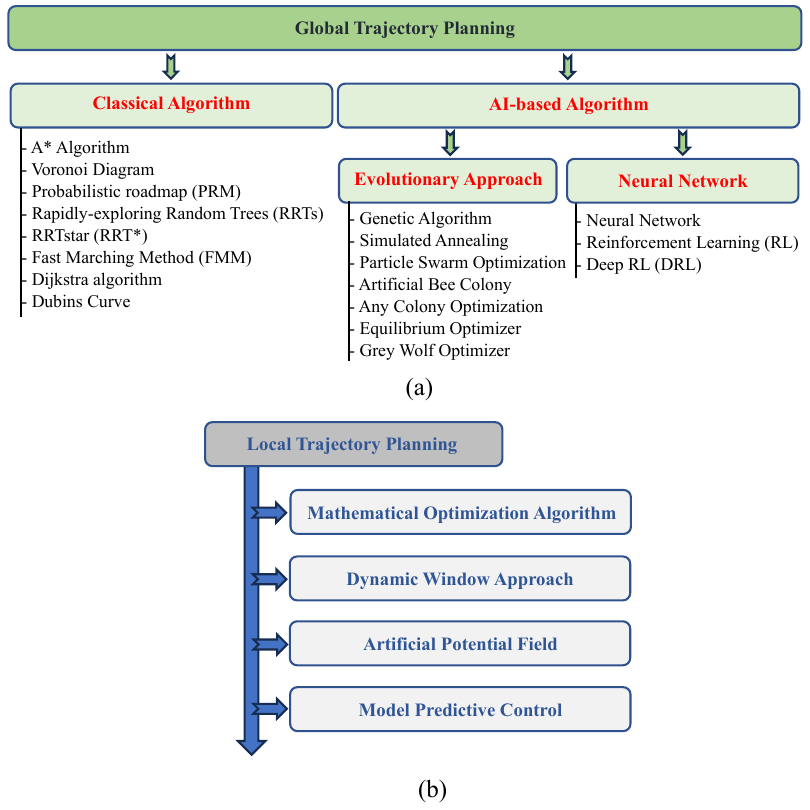}
		\par\end{centering}
	\caption{\label{fig:Planning}Illustration of common trajectory planning techniques.}
\end{figure}

Trajectory planning can be broadly categorized into two main types:
global trajectory planning and local trajectory planning. Global trajectory
planning typically uses either classical algorithms or AI-based methods.
Classical algorithms include A-star (A{*}), Voronoi Diagrams, Probabilistic
Roadmaps (PRM), Rapidly-exploring Random Trees (RRT), RRTstar (RRT{*}),
the Fast Marching Method (FMM), Dijkstra's Algorithm, and Dubins Curves.
AI-based approaches encompass evolutionary algorithms and neural networks.
Examples of evolutionary algorithms are Genetic Algorithms (GA), Simulated
Annealing, Particle Swarm Optimization (PSO), Artificial Bee Colony
(ABC), Ant Colony Optimization (ACO), the Equilibrium Optimizer (EO),
and the Grey Wolf Optimizer (GWO). Neural network techniques include
artificial neural networks, Reinforcement Learning (RL), and Deep
Reinforcement Learning (DRL). In contrast, local trajectory planning
focuses on real-time adaptability using mathematical optimization
techniques, the Dynamic Window Approach (DWA), Model Predictive algorithm,
and Artificial Potential Fields (APF). Fig. \ref{fig:Planning} illustrates
the taxonomy of these common trajectory planning methods.

Cell decomposition techniques result in an unbounded range of movement
for UAVs resulting in a very large search space. Roadmap approaches
consist in creating a connectivity graph built of linked nodes that
represent key unoccupied space \cite{chen2021path}. Popular roadmap
approaches include visibility graph, Voronoi diagram, PRM, RRTs, RRT{*},
A{*}, and evolutionary techniques \cite{kumar2019lego}. Visibility
graph consists in forming a graph representation of all potential
visible connections in the environment which is tedious in the construction
process \cite{niu2019voronoi}. Voronoi diagram offers an alternate
faster approach than visibility graph as it decomposes the space into
a set of polygon regions such that each region is formed around a
single environment \cite{niu2019voronoi}. PRM is a probabilistic
approach that deconstructs the space into a group of randomly placed
connectivity nodes where environment knowledge is needed in the process
of path construction \cite{chen2021path}. Similar to the PRM, RRTs
approach requires highly detailed knowledge of the environment to
design an explorative branching strategy originating from a root node
to the target destination \cite{chen2021path}. Artificial intelligence
evolutionary techniques can also be utilized for path planning and
examples include ACO (e.g., salesman problem), PSO, ABC, and Optimization
Equilibrium algorithms \cite{orozco2019mobile,yu2022novel,xu2020new}.
Both cell decomposition and roadmap techniques construct an environment
representation based on prior knowledge of the environment. Finally,
APF method focuses on calculating directional attractive forces in
the direction of the target location, whereas obstacles generate repulsive
forces. APF techniques are computationally efficient and allow the
UAV to travel accurately and quickly to the desired destination.

\subsection{Collision Avoidance}

CA techniques are typically incorporated into the planning algorithms
either using a reactive or a deliberative approach \cite{prasad2020geometric,nister2019safety,he2020optimisation,ramasamy2016lidar}.
The reactive approach relies on real-time information about the surroundings
gathered by local on-board sensors allowing for fast response to sudden
environment changes with the limitation of possible stuck in local
minima. Deliberative planning, on the other hand, requires an accurate
updated map of the environment to perform the path planning calculation
and identify the optimal collision free route to the target location.
Consequently, deliberative planning is challenged with higher computational
power needs and unsuitability for dynamic environments. Perception
is an indispensable step towards obstacle detection and forms the
basis for CA approaches. Popular perception and CA sensors include
stereo or monocular cameras, thermal or infrared cameras, infrared
devices, ultrasonic systems, LiDAR, solid-state systems, or optomechanical
devices. For commercial aircraft, TCAS offers resolution advisories,
however it is sensitive to jamming as will be discussed later. A UAV
can use TCAS in conjunction with ADS-B to achieve CA \cite{chand2017sense}.
Once obstacles have been located, there are three popular control
maneuver actions can be applied to avoid a conflict. These actions
that can be applied alone or in combination and include ``climb/descent''
(altitude control maneuver), ``turn right/left'' (heading control
maneuver), and acceleration/deceleration (speed control maneuver)
\cite{chand2017sense}. Popular CA algorithms are as follows: geometric
approach (relies on distance between UAV and obstacle) \cite{prasad2020geometric},
force-field (relies on attractive (final destination) or repulsive
forces (obstacles)) \cite{nister2019safety}, optimization-based \cite{he2020optimisation},
sense-and-avoid (deviates the UAV from its original route and return
it back to reduce computation cost) \cite{ramasamy2016lidar}.

\subsection{Trajectory Control}

Trajectory tracking control executes the planned path. One of the
major challenges of controlling a UAV is the fact that in most cases
its an underactuated system. For example, quadrotor UAVs have a total
of six outputs (three rotational angles (Roll-Pitch-Yaw angles) and
translational position in x-y-z axes) but only total of four inputs
(thrust and three rotational torques) \cite{hashim2022ExpVTOL}. Thus,
the strong interdependence between UAV position and attitude overcomplicates
the control system design. To address the underactuation issue, typical
controllers are designed in a cascaded manner with an outer and an
inner control \cite{hashim2022ExpVTOL,hashim2023_ISA}. The outer
control receives the desired (reference) and actual UAV positions
and based on the position error generates the required thrust and
the desired UAV orientation (rotational angles). The inner control
receives the desired and actual UAV orientation and generates the
required rotational torques. Given the required thrust and rotational
torques, the necessary rotor speed can be calculated \cite{hashim2022ExpVTOL,hashim2023_ISA}.
Note that to ensure UAV stability, the inner control (attitude control)
must be designed faster in terms of control and response than the
outer control (position control) \cite{hashim2022ExpVTOL}. Fig. \ref{fig:EstimatorControl}.(b)
presents classification of trajectory control design. Common controllers
include linear control \cite{kamel2017dynamic}, optimal control \cite{arifianto2015optimal},
feedback linearization \cite{ali2024mpc}, Lyapunov-based \cite{hashim2022ExpVTOL,hashim2023_ISA},
backstepping \cite{shevidi2024quaternion}, Sliding Mode Control (SMC)
\cite{shevidi2024quaternion}, Model Predictive Control (MPC) \cite{ali2024mpc},
Fuzzy Logic Control (FLC) and Adaptive Fuzzy Neural (AFN) \cite{rao2022position},
and learning based approaches (Lyapunov-based Adaptive Neural Control
(LyANC) or reinforcement learning (RL)) \cite{ali2024deep,hashim2022NNCont}.
The UAV model is highly nonlinear and is typically described with
respect to the Special Euclidean Group $SE(3)$. Hence, the linear
control design relies on linearizing the UAV model around multiple
nominal points where each point represents certain UAV pose configuration.
Afterwards, a gain scheduling approach that switch between different
proportional-integral-derivative PID controllers is designed such
that the PID control parameters change with respect to the linearized
UAV model at its current nominal point \cite{shao2020robust}. The
gain scheduling-based PID controller is a standard for tracking control
of UAVs and airplanes due to the simplicity of its design and implementation.
However, it is not suitable for fast maneuvers.

Optimal control minimizes control inputs which, in turn, reduces the
energy used by rotors as well as limits the tracking trajectory error.
Optimal control approaches include Linear Quadratic Regulator (LQR)
and Linear Quadratic Gaussian (LQG) \cite{arifianto2015optimal}.
Feedback Linearization (FBL) starts with formulating the UAV model
in a state-space form. It is worth noting that for FBL the model has
to be restructured into a fully-actuated form which is typically satisfied
around nominal points. The advanced version of FBL can be merged with
a Control Barrier Function (CBF) to enhance the tracking control while
ensuring collision avoidance \cite{ali2024mpc}. The nonlinear family
includes such controllers as Lyapunov-based, backstepping, and SMC
which use the UAV nonlinear model directly \cite{hashim2022ExpVTOL,hashim2023_ISA}
allowing for fast maneuvers. By extracting the nonlinear control algorithm
from a Lyapunov candidate function it becomes possible to guarantee
error reduction toward zero or to the neighborhood of the origin \cite{hashim2022ExpVTOL,hashim2023_ISA}.
Moreover, tracking control algorithms can be combined for optimal
performance. For instance, its common to merge backstepping control
with SMC-type controllers \cite{shevidi2024quaternion} or with FBL.
SMC subfamily consists of classic SMC \cite{bingol2021neuro}, fast
terminal SMC (FTSMC) \cite{shevidi2024quaternion}, nonsingular terminal
SMC (NTSMC) \cite{labbadi2022barrier}, fractional terminal SMC SMC
(FrTSMC) \cite{yu2018distributed}, and super twisting SMC (STSMC)
\cite{babaei2019adaptive} where each type has its own characteristics.
MPC is an optimization-based control approach also known as a receding
horizon control since the prediction horizon is shifted over the processing
time \cite{ali2024mpc}. MPC is extracted from a custom-built objective
function that penalizes components, such as tracking error, control
effort, uncertainties, and disturbances, among others \cite{ali2024mpc}.
Although MPC is able to handle constraints on states and inputs, its
shortcoming is higher computational cost when compared to the above-mentioned
controllers. Finally, learning-based controllers are model free and
they operate using either Lyapunov-based neural adaptive or reinforcement
learning \cite{ali2024deep,hashim2022NNCont}. Learning-based controllers
utilize an approximated UAV model that is tuned iteratively until
reasonable performance is achieved. 

\begin{figure*}
	\centering{}\includegraphics[scale=0.53]{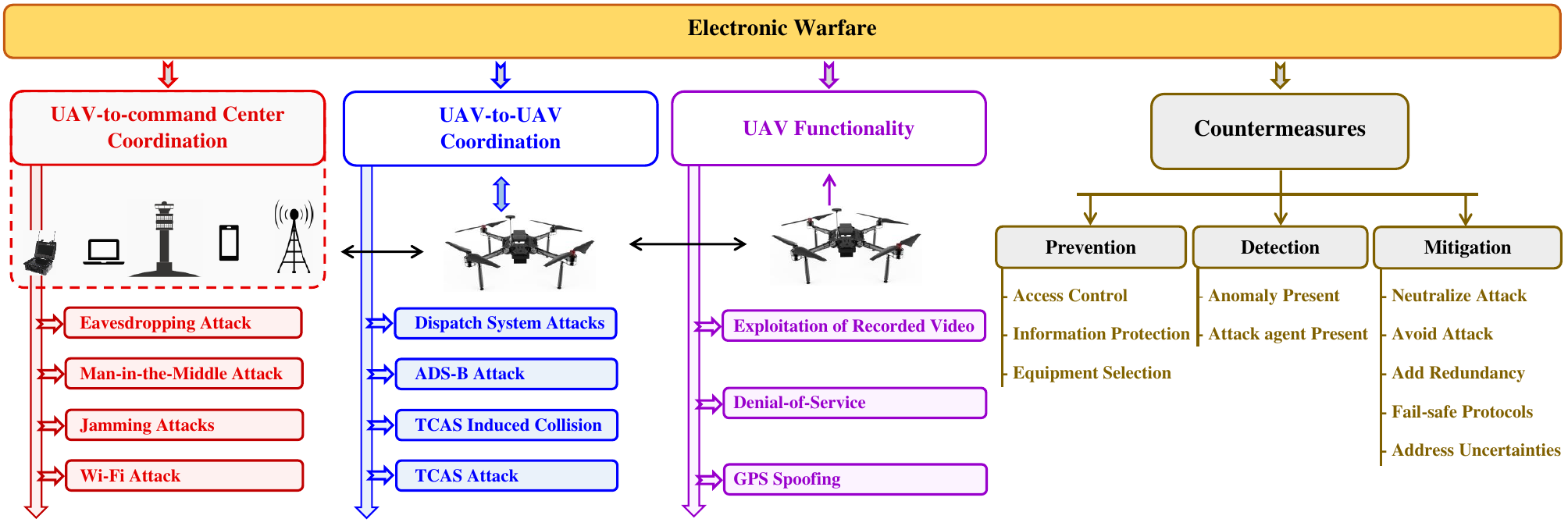}\caption{Taxonomy of UAV electronic warfare attacks and countermeasures.}
	\label{fig:UAV_EW2}
\end{figure*}

\section{Electronic Warfare and UAVs\label{sec:ElectronicWarfare}}

Warfare has long been studied by adversaries to assess the strengths
and weaknesses of their opponents, with the aim of minimizing risks.
Key factors on the battlefield include military planning, which ensures
communication between all forces, both on the battlefield and at staff
headquarters; air defense, which utilizes radar to detect incoming
airborne threats and processes intelligence received through landlines
or data links; air-superiority aircraft, which employ quick reaction
alerts to disrupt or slow enemy air patrols; defense suppression,
which uses radar for terrain-following and target acquisition; and
offensive operations, which rely on radar to locate targets and guide
missiles \cite{moir2019military}. The electromagnetic (EM) spectrum,
along with directed energy, constitutes a fundamental component of
electronic warfare. Electronic warfare can therefore be defined as
the science of maintaining friendly use of the EM spectrum while denying
its use to the enemy. Specifically, it seeks to control the EM spectrum,
potentially disrupting or blocking enemy radar signals and data communications,
thereby rendering their systems inoperative. The research community
is actively working on developing advanced strategies and techniques
that leverage the EM spectrum to protect friendly UAVs from adversarial
threats while disrupting or neutralizing hostile UAV operations \cite{moir2019military,cevik2013small}.
The rapid growth of digitally-operated aerial vehicles further positions
cyberattacks as a crucial domain within electronic warfare. Common
electronic warfare cyberattack threats targeting UAVs can be categorized
into three main areas: UAV-to-command center coordination, UAV-to-UAV
coordination, and UAV functionality. UAV-to-command center coordination
threats include eavesdropping attacks, Man-in-the-Middle (MITM) attacks,
jamming attacks, and Wi-Fi-based attacks. UAV-to-UAV coordination
threats involve dispatch system attacks, ADS-B (Automatic Dependent
Surveillance-Broadcast) attacks, TCAS (Traffic Collision Avoidance
System) induced collisions, and TCAS-specific attacks. UAV functionality
threats typically focus on the exploitation of recorded video feeds,
Denial-of-Service (DoS) attacks, and GPS spoofing. Countermeasures
against these threats fall into three categories: prevention, detection,
and mitigation. Prevention countermeasures involves measures such
as access control, information protection, and careful equipment selection.
Detection countermeasures relies on identifying anomalies or detecting
the presence of attack agents. Mitigation countermeasure strategies
include neutralizing or avoiding the attack, adding redundancy, implementing
fail-safe protocols, and addressing uncertainties. For more details
visit \cite{Yu2024EW}. Fig. \ref{fig:UAV_EW2} illustrates the taxonomy
of these common UAV cyberattack threats and countermeasures.

\subsection{Destructive and Non-destructive Cyberattacks}

\begin{figure*}
	\centering{}\includegraphics[scale=0.88]{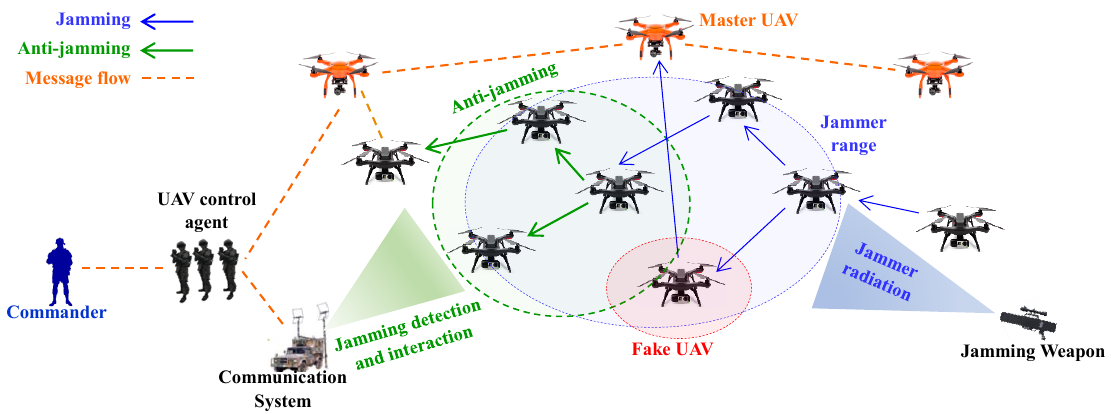}\caption{UAVs in the middle of electronic warfare illustrating spoofing, MITM,
		jamming attacks, and anti-jamming reactions.}
	\label{fig:UAV_EW}
\end{figure*}

Cyberattacks can be broadly classified as destructive and non-destructive
\cite{kratky2017non}. Non-destructive cyberattack approaches refers
to the situation when there is no direct destruction of the affected
system. For instance, non-destructive cyberattack methods can be seen
in a form of protected information leakage, use of unlawful information,
integrity violation, and/or denial of a service. Leakage of protected
information can occur when UAV downlink channel information is accessed
\cite{wu2019safeguarding}. An example of unlawful information use
includes accessing non-public services without lawful authorization
\cite{wu2019safeguarding}. Destroying or changing UAV's data represent
an incident of integrity violation \cite{bramlette2019cyber}. Non-destructive
cyberattacks can be done via Denial-of-Service (DoS) attacks or Distributed
DoS (DDoS) attacks through disabling real-time activities by systematically
sending requests. Further examples of non-destructive cyberattacks
include malware, an active attack (e.g., MITM), Address Resolution
Protocol (ARP) spoofing, multi-layered jamming, blockhole, and deauthentication
attacks \cite{seo2022d}. MITM attacks secretly intervene into multi-UAV
system communication links penetrating the communication channel between
the legitimate sender and the receiver to steal or modify information
packets. Spoofing attacks bypasses the predefined access control rules
of communication network in response to a specific target request
\cite{huang2018combating}. In other words, spoofing occurs when an
unknown unsafe source of communication is disguised as a known trusted
source. In case of GPS spoofing, forged location information is entered
into a program to produce false signals, which in turn are broadcast
with Universal Software Radio Peripheral (USRP) \cite{meng2021approach}.
As a result, UAV receives false location and time information invalidating
all of its GPS functions. Multi-layered jamming targets radio communication
and disturbs information transmission and reception. Blockhole attack
is a routing cyber warfare threat, hat presents itself as a route
node, projects false optimized real-time routing path to the target
node (in response to a request), and instead of relaying information
packets, absorbs them. De-authentication attack consists in the adversary
transmitting de-authentication packets to interrupt or disconnect
the target UAV connection and gain control over the system. 

Destructive cyberattack methods focus on lower layers of the Open
Systems Inter-connection (OSI) mainly targeting individual hardware
elements necessary to the use of multiple systems and mechanical systems
\cite{dahiya2020unmanned}. Moreover, destructive cyberattack can
focus on overheating the UAV battery system \cite{dahiya2020unmanned}.
UAVs heavily rely on wireless communication, and as a result, the
most common electronic warfare threats for UAVs include service disruption,
hijacking, data integrity damage, and remote code execution \cite{ghosh2020uav}.
Besides, GCSs can be targeted as part of the electronic warfare. Fig.
\ref{fig:UAV_EW} illustrates multiple electronic warfare attack techniques
and anti-jamming mitigation. Furthermore, malware is a dangerous tool
of electronic warfare. Malware, software designed to cause disruption,
can infect one or more UAVs by spreading through a networked group
or arriving from a GCS with the objective of taking control over the
UAVs. Another common destructive threat are the Hardware Trojans that
focus on hardware level malicious modifications to the flight controller
circuit \cite{asif2024confide}. The underlying systems that comprise
the flight controller are complex, and thereby the Trojans are commonly
used. Introduction of the imitation hardware to the supply chain forms
a security breach of the circuit itself resulting in permanent failure
of components potentially leading to UAV destruction \cite{asif2024confide}.

\subsection{Attacking Transponders}

Identification systems discussed in Section \ref{sec:IdentificationSystem}
are crucial for the safe operation of UAVs and transponders are their
essential component. Due to their importance, UAV transponders are
an key target of electronic warfare. ADS-B can be attacked through
unprotected messages transmitted in plain text format \cite{wu2020security}
with no authentication methods used to prevent message tampering \cite{zhang2021design}.
This type of attack can be classified as message elimination, spoofing-based
message infusion, and message fabrication \cite{zhang2021design}.
Message elimination utilizes external transmitters to introduce constructive
or destructive interference into the ADS-B signal. Constructive interference
involves introduction of bit errors into the ADS-B message causing
the UAV's receiver to disregard the detected messages, and therefore,
weakening the UAV's transmitter awareness \cite{wu2020security}.
On the other side, destructive interference involves sending an attacking
signal which constitutes an inverse of the original ADS-B signal causing
complete or partial destruction of the message \cite{wu2020security}.
Spoofing-based message infusion focuses on injecting malicious messages
into the airspace, causing ADS-B receivers to perceive appearance
of illegitimate aircraft. Note that ADS-B does not use any authentication
methods in messages. Hence, message infusion can be applied using
commercially available devices. Furthermore, the broadcast false messages
could target either the UAVs themselves which is commonly known as
``aircraft target ghost injection'' or target the ground command
center which is termed ``command center ghost injection''. The targeted
ADS-B receiver will see an illusive aerial vehicle in the airspace
as the attacker anonymously manipulates the air traffic \cite{yang2018practical}.
Message fabrication represents manipulation of ADS-B signals by introducing
false information which ADS-B receivers regard it as interpret. In
contrast to message infusion, message fabrication manipulates real
messages transmitted by legitimate UAVs. Consequently, the level of
tampering varies in accordance with the intention. The attacker has
to broadcast a very high-powered ADS-B signal in order to substitute
parts or the whole of the ADS-B message, which is commonly known as
overshadowing \cite{yang2018practical}. All of the three types of
ADS-B message attacks pose a serious threat since they can allow enemies
to masquerade as potential allies.

TCAS resolution advisories do not predict long term effects resulting
in a challenging issue commonly known as TCAS induced collision, where
the suggested resolution advisories presented by TCAS could cause
a collision \cite{weinert2022near}. Moreover, TCAS is known to pose
a safety concern as it is not designed to withstand cyberattacks (e.g.,
jamming \cite{habler2023assessing}). The bad actors could take advantage
of this weakness and jam the 1090 MHz channel to stop the aerial vehicle
from tracking potential intruders. Nonetheless, these attacks can
be spotted and addressed \cite{habler2023assessing}. Popular and
challenging to counteract jamming attacks are known as “All-Call Flood”
and “Squitter Flood”. The “All-Call Flood” attacks leverage All-call
interrogation and use the 1030 MHz channel to trigger all nearby Mode
S transponders to respond with their 24-bit International Civil Aviation
Organization (ICAO) code flooding1090 MHz reply channel \cite{habler2023assessing}.
The “Squitter Flood” attack represents spoofing actions to the nearby
transponders in the form of transmitted replies on the 1090 MHz channel
which forces the transponders to continuously track a “false” aerial
vehicle \cite{mohen2021cybersecurity}. These two types of attacks
increase the chances of Near Midair Collision (NMACo) events. Nonetheless,
the attacker might not gain full control over the NMACo occurrence.
Other types of transponder attacks unrelated to channel flooding are
termed “Phantom Aircraft” attacks. In these cases the attacker may
succeed to generate accurate Mode S replies and to be perceived as
a moving airplane by the TCAS transponders \cite{weinert2022near}.
Consequently, the TCAS will assume that these replies are received
from an actual aircraft and will be forced into tracking it.

\subsection{Jamming and Anti-jamming}

UAVs can be subject to constant jamming attacks \cite{habler2023assessing},
where malicious equipment is dedicated to broadcasting a continuous
high power interference signal occupying the channel and impeding
the legitimate UAV’s packet reception \cite{darsena2021detection}.
Reactive jamming attacks are energy efficient techniques since malicious
signal is sent only when transmission of legitimate data packets is
detected \cite{darsena2021detection}. Deceptive, random and periodic
jamming attacks are commonly employed but are less effective \cite{alrefaei2022survey}.
Man-in-the-middle attacks and spoofing are typical attacks that affect
reconnaissance data confidentiality and integrity with the aim of
causing loss of wireless connection for vehicle-to-vehicle (V2V) and
vehicle-to-infrastructure (V2I) communications in vehicular ad-hoc
networks (VANETs) \cite{pirayesh2022jamming}. In VANETs, counteracting
jamming attacks are critically important, since any connection loss
could potentially lead to vehicle collision or loss. This may not
only result in mission failure but also pose a public safety risk.
When targeting wireless connection loss in V2V and V2I communications
in VANETs, jamming attackers could utilize a variety of strategies,
such as constant, reactive, and deceptive jamming discussed above
\cite{ren2024k}%
. Therefore, anti-jamming mitigation actions are an essential component
of electronic warfare. Jamming attacks can be minimized using Packet
Delivery Ration (PDR) and RSS metrics in application to master and
slave UAVs. Furthermore, anti-jamming can employ classical wireless
techniques, such as channel hopping and spectrum spreading \cite{darsena2021detection}.
The anti-jamming techniques used depend on the type of jamming they
aim to counteract. For instance, power control game modeling technique
is applicable for constant jamming \cite{lv2017anti}, Bayesian Stackelberg
game modeling \cite{xu2019joint}, and learning based frequency which
is more applicable to multi-agent UAVs \cite{peng2019anti}. Synchronization,
modulation, demodulation, channel equalization and estimation are
key factors that must be considered when investigating jamming and
anti-jamming methods.

\subsection{Countermeasures and Defensive Aids}

Countermeasures and defensive aids is the science of protecting information
handled by UAVs with regard to confidentiality, authenticity, and
integrity in addition to ensuring service availability for each of
civilian and military applications \cite{nguyen2021drone}. Based
on different types of attacks to UAVs discussed above, countermeasures
strategies to these attacks are involved to respond to gaining control
over the spectrum, software issue, vulnerabilities found on hardware,
and network layers. Countermeasure are broadly classified into three
categories: prevention, detection, and mitigation \cite{moir2019military}.
Prevention can be achieved through the following three methods: access
control, information protection, and component selection. Access control
ensures that the UAV can receive communication only from authorized
personnel or authorized software via password-based node authentication
schemes (e.g., Media Access Control (MAC) address) \cite{khan2021dual}.
Information protection focuses on message interception, elimination,
or infusion (e.g., cryptography) \cite{khan2021dual}. Component selection
involves anti-tampering technologies employed onboard of the UAVs
to prevent entry points from getting potential attacks \cite{kong2021survey}.
If the prevention methods failed, detection countermeasure is applied
and the popular approach is presence of anomaly (detecting abnormal
patterns such as radio signals, communication traffic, and/or flight
behavior) \cite{kong2021survey}. Mitigation is applied to overcome
the attack through neutralize/avoid, redundancy (e.g., switching the
GNSS constellation), and fail-safe (e.g., UAV returns to the home
base or self-destruct in case of connection loss).

\section{Databus in UAVs\label{sec:Databus}}

Efficient data transfer and communication are critically important
to UAV operation which relies on real-time data transfer and databus
facilitates this process. Modern databuses replace point-to-point
wiring with centralized and streamlined connections \cite{kornecki5avionics}.
However, majority of aircrafts are still using point-to-point wiring
databuses (e.g., Aeronautical Radio Incorporated (ARINC) 429) introduced
in 1960s which simply connect components together using individual
wires. Common bus topologies used with ARINC 429 include star topology,
bus-drop topology, and multiple bus topology \cite{moir2013civil}.
Star topology involve a single Line Replaceable Unit (LRU) transmitter
that sends data to one or more different receivers. Bus-drop topology
incorporates multiple receivers positioned along a physical path.
However, the bus-drop topology faces several challenges related to
aerial vehicles weight and maintenance due to bulky onboard cabling
\cite{safwat2014evolution}. Faster and lighter bi-directional databuses
present an alternative solution. In case of a two-way LRU's communication,
multiple bus design is typically utilized. The military standard MIL-STD-1553
was developed as a bi-directional data communication bus system for
avionics applications to address the increasing complexity of point-to-point
wiring in avionics systems \cite{de2021exploiting}. MIL-STD-1553
incorporates the following major components: transmission media (databus),
remote terminals (RTs), bus controllers (BCs), and bus monitors (BMs)
\cite{de2021exploiting}. A typical RT is composed of an encoder/decoder
transceiver, transceivers, a buffer or memory, a protocol controller,
and a subsystem interface. BCs manage and control data-flow through
the databus by sending commands to RTs \cite{spitzer2017digital}.
BMs, on the other hand, are tasked with monitoring the data transmission,
maintenance, and flight-test recording and BMs can be used as a backup
system for the BC. MIL-STD-1553 provides reliable operation in the
harsh physical and electromagnetic environments for both military
and commercial applications. MIL-STD-1553 has been deployed on fixed
and rotary wing aircraft, ground vehicles, spacecrafts, satellites,
and unmanned aircraft \cite{spitzer2017digital}. 

Controlled Area Network (CAN) is a standard UAV databus that facilitates
communication between sensors, cameras, actuators, the main controller.
CAN bus is a message-based protocol that enables Electronic Control
Units (ECUs) to communicate with each other by using a priority mechanism
\cite{lin2005reliability} through node communication. With the rise
of automobile industry, CAN bus was born out of a collaboration between
Bosch, Intel, and Mercedes Benz in 1986 to promote faster communication
between the large number of ECUs present in a car \cite{lin2005reliability}.
Bosch published CAN 2.0 around 1991 which was standardized (International
Organization for Standardization (ISO) 11898) around 1993. Later in
2003, the data link layer was separated from the physical layers,
and in 2015 Flexible Data-Rate (CAN FD) was developed. As opposed
to previous databus standards, CAN FD enables more flexible and greater
volume data transfer at a higher speed ranging between 1 an 8 Mbps.
CAN FD permits 64 bits messages in contrast to 8 bits short messages
used by its predecessor. This enables the ECU’s to decide on the message
size as well as dynamically change their data transmission speed \cite{pohren2019analysis}.
A typical data transfer involves the following steps: decoding, start-of-frame
(node is intended to talk), standard identifier (refers to priority
level), remote-transmission request (whether to talk to another node
or not), control commands, actual data message, error detection and
correction, acknowledgment (data received correctly), cyclic redundancy
(ensure data integrity), and end-of-frame. Nowadays, a variety of
databuses and data transmission protocols are used in UAV applications,
such as CAN Bus, MIL-STD-1553, ARINC 429, ARINC 814, universal asynchronous
receiver/transmitter (UART), Recommended Standard (RS) serial communication
RS-232, Ethernet, Avionics Full-Duplex Switched Ethernet (AFDX), Fiber
optics, Bluetooth, IEEE 802.15.3, Wireless LAN, and many others. 

In terms of redundancy, ARINC-429 does not inherently provide a redundancy
feature. MIL-STD-1553, on the other hand, incorporates built-in redundancy
by including two built-in twisted shielded pairs in addition to a
monitor which automatically activates when the bus controller is down
\cite{gangl2013case}. Although CAN bus does not have any built-in
redundancy components, it uses the cyclic redundancy check (CRC) to
detect errors in case an issue arises. Signal reliability and robustness
are critical requirements in harsh electromagnetic environments. With
regard to signal interference (e.g., electromagnetic waves), MIL-STD-1553
has higher immunity when compared to CAN Bus and ARINC \cite{bracknell2007mil}.
As to encoding/decoding, ARINC-429 uses bi-phase modulation often
referred to as bipolar return to zero (BRZ), whereas MILSTD-1553 relies
on Manchester encoding using Non-Return-to-Zero (NRZ) encoding \cite{spitzer2017digital}.
Likewise, CAN bus and UART use NRZ encoding where low-to-high transition
represents a 1, and a high-to-low refers to a 0, similar to Manchester.
Considering day-to-day avionics advancements, legacy protocols ARINC-429
and MIL-STD-1553 are insufficient to address the increased bandwidth
demands. Nonetheless, their legacy networks are highly reliable for
the current applications. Although the CAN bus can provide high data
rates, it is not regarded as reliable as ARINC-429 and MIL-STD-1553.

\section{Regulations, Safety, Classification, and Certification \label{sec:RegulationsCertification}}

\subsection{Failure Classification and Design Philosophies}

Regulations and reliability of UAV solutions must keep pace with rapid
growth of UAV industry to guarantee continued success and public safety.
Therefore, all potential sources of accidents should be identified
and the probability of failure minimized through redundant and/or
fail-safe UAV avionics systems made up of reliable components. Failure
types can be classified into four groups: catastrophic, hazardous,
major, and minor. Catastrophic classification is the event that prevent
continued safe flight and/or landing and the expected consequence
is multifatal accident (e.g., death). The probability of occurrence
of a catastrophic failure is $1\times10^{-9}$ FH (per flight hour),
in other words, such incidents should be extremely improbable. Hazardous
failure could result in a serious incident (some injuries or loss
of life) and its probability is $1\times10^{-7}$ FH. Major failure
conditions, on the other hand, may lead to difficulties, but the aircraft
will be able to fly and land safely, and its probability is $1\times10^{-5}$
FH. Minor failure implies that the UAV intelligent systems and/or
the human operator will be able to take the necessary mitigating actions
and its probability should not be greater than $1\times10^{-3}$ FH.
Design of different UAV avionics systems must carefully account for
the above-discussed failure classification probabilities. The avionics
systems are typically created following one of the two design philosophies:
Safe-Life or Fail-Safe. The Safe-Life philosophy designs each component
or structure to operate free from failure during its lifetime accurately
estimated through research analysis. By contrast, the Fail-Safe philosophy
has the assumption that failure will eventually occur and it incorporates
various techniques to handle losses based on system/component failures
in a safe manner.

\subsection{Incidents and Safety}

\begin{figure}
	\centering{}\includegraphics[scale=0.2]{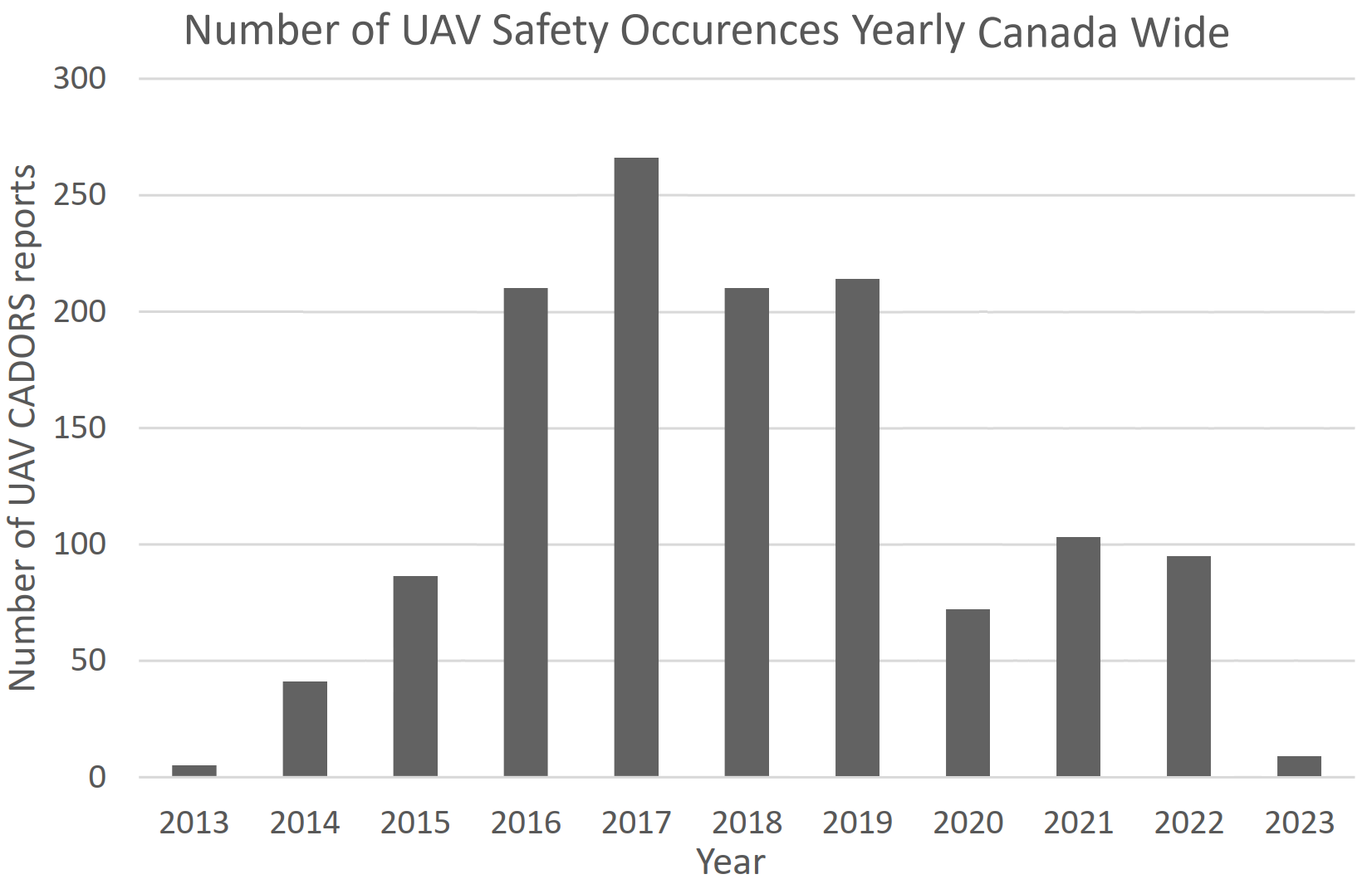}\caption{Total number of UAV Safety Occurrences in Canada between 2013 and
		2023 \cite{TransportCanada}.}
	\label{fig:UAV_incidents}
\end{figure}

The International Air Transport Association (IATA) documented an increase
of UAVs safety risks caused by unpredictable nature of UAV operators
and inability to determine their location \cite{wanner2024uav,RiskWerfelman}.
The IATA’s safety report \cite{RiskWerfelman} also illustrated that
50\% of UAV related incidents between 2014 and 2018 were documented
in Europe. However, in recent years the European Aviation Safety Agency
(EASA) reported a significant drop in UAV safety occurrences (from
minor to catastrophic) \cite{wanner2024uav,EASA2022}. Furthermore,
in 2021, severe incidents, defined as occurrences resulting in injury
or fatalities, were near zero \cite{wanner2024uav,EASA2022}. EASA
attributed the drop in UAV safety occurrences to the introduction
of UAV pilot certification requirements and technology advancement
(e.g., obstacle detection and avoidance). In line with IATA and EASA
reports, in recent years, Canada has seen a significant improvement
in terms of the number of recorded UAV safety occurrences. Using the
Civil Aviation Daily Occurrence Reporting System (CADORS) database
from Transport Canada Civil Aviation (TCCA), the aviation safety occurrences
involving UAVs were recorded between January 2013 and March 2023.
Fig. \ref{fig:UAV_incidents} illustrates the annual number of Canadian
UAV safety occurrences where pilots of civil and military aircraft
spotted UAVs entering their flight path. The data reveals that the
Canadian UAV safety occurrences reached their peak in 2017, the same
year that an incident was reported at Quebec City’s Jean Lesage airport
involving a collision between a small drone and a passenger airplane
causing minor damage. In spite of the fact that the incident caused
only minor damage, the potential for a catastrophic outcome was recognized
leading to major revisions and new interim rules and regulations by
Transport Canada. As a result, to date, no sever safety occurrences
involving UAVs have been recorded in Canada \cite{TransportCanada}.

\subsection{Airspace Classification}

A common cause for safety incidents is a UAV entering a class of airspace
that they are not permitted to fly in. For instance, Canada subdivides
the airspace into seven classes (A, B, C, D, E, F, and G) where each
class has its own rules (e.g., type of aircraft and communication
equipment) \cite{wanner2024uav}. NAV CANADA provides ATC and flight
information to every class of the airspace to ensure air safety. Class
A and B are referred to as controlled high-level (over 18,000 ft)
and low-level (12,000 to 18,000 ft) airspace, respectively. Class
A and B involve pilot clearance signals and instructions for maintaining
safe distance from other aircraft. Class C, D, and E refer to terminal
areas for busy airports, control zones and terminal areas for moderate
traffic airports, and control zone for airports without towers, respectively.
Class F describes a restricted and advisory airspace (no aircraft
is allowed to enter without a permission from the controlling agency).
Class G represents uncontrolled airspace and it reaches an altitude
of 18,000 ft \cite{wanner2024uav,grote2022sharing}. In Canada, UAVs
must complete their activities in within Class G airspace as well
as obey rules put in place by Transport Canada. A UAV can fly in Class
A to E airspace only provided that (i) it is registered with Transport
Canada, (ii) the UAV operator holds a Pilot Certificate - Advanced
Operations, and (iii) the operator has a written authorization from
NAV CANADA \cite{NavCanada}. Unfortunately, UAV safety occurrences
tend to take place when UAV operators do not follow the above-listed
rules.

\subsection{Certification}

A pilot operator wishing to fly a UAV weighing between 250 gm and
25 kg must possess a pilot certificate under the Canadian regulations
901.54 and 901.62 \cite{wanner2024uav}. Transport Canada defines
two UAV certificates: Pilot Certificate Basic Operations and Pilot
Certificate Advanced Operations where one or both them can be obtained
through an online exam administered by Transport Canada. Pilot Certificate
Advanced Operations requires the examinee to demonstrate a much higher
understanding of concepts similar to other forms of transportation
exams (e.g., automobiles) \cite{TransportCanada2}. Examinations for
both basic and advanced certificates cover almost the entirety of
Canadian Aviation Regulations (CARs), General Provisions, and General
Operating Flight Rules. Candidates must demonstrate proper knowledge
of Piloted Aircraft Systems (RPAS) airframes, powerplants, and propulsion
and systems, and ability to handle and secure electrical systems,
launch and recovery systems, datalinks, batteries, autopilots, electrical
motors, payloads, ground control stations, redundancies and critical
items, and other systems onboard UAVs \cite{TransportCanada2}. Furthermore,
candidates must demonstrate understanding of other crucial concepts,
such as air traffic rules, air law, human factors, navigation, flight
theory, radiotelephony, procedures, flight operations, and meteorology
\cite{TransportCanada2,moir2013civil}. Both basic and advanced pilot
certifications require the operators to register their UAVs with Transport
Canada in compliance with regulation 901.06 using the Transport Canada’s
Drone Management Portal \cite{TransportCanada3}.

\section{Persisting Challenges and Future Perspectives}

\paragraph*{Autonomy}Deploying a single UAV or a fleet of UAVs to
autonomously perform complex tasks without human intervention may
currently seem like a utopian concept. While this level of autonomy
has not yet been fully realized, significant advancements in the field
indicate that this vision could become a reality in the near future.
Constrained and predictive adaptive real-time controllers and estimators
(e.g., navigation, SLAM, and pose estimation), which are pillar components
for achieving full autonomy, require substantial development efforts
to ensure system stability, efficient trajectory planning, and energy
optimization. These advancements contribute to maximizing flight duration
and enhancing payload capacity. Moreover, managing the heavy real-time
data streams associated with autonomous UAV operations necessitates
integrating advanced Artificial Intelligence (AI) techniques such
as machine learning (including supervised, unsupervised, neural adaptive,
and reinforcement learning), evolutionary algorithms (for single and
multi-objective optimization), and fuzzy logic expert systems. These
techniques are essential for enhancing decision-making in autonomous
missions, e.g., trajectory planning, navigation and control, battery/energy
scheduling, obstacle avoidance, detection and classification, and
resource allocation. Processing large volumes of real-time data using
diverse decision-making algorithms (e.g., path planning layer, control
layer, communication layer, obstacle avoidance layer, and perception
layer) requires high-speed onboard processing units with substantial
memory capacity to support these complex computations.

\paragraph*{Advanced Technology}Future efforts should focus on developing
advanced, high-quality, and higher-speed but compact and low-cost
sensing devices, including LiDAR, infrared, thermal cameras, specialized
cameras, multispectral imaging, RFID, UWB, bluetooth low energy, IMU,
magnetic, multi-decision-making algorithms, UAVs antennas and integrating
edge/fog computing systems. Additionally, exploring hybrid power supply
systems (such as fuel cells, solar cells, and batteries) and solar
energy harvesting will be crucial for maximizing UAV flight duration
effectively. There is a growing interest in integrating cellular networks
with UAV networks to support future wireless technologies. Research
opportunities include efficient handovers for mobile UAVs and joint
sensing and communication for mmWave and UWB sensors. While initial
simulations show promise, this field is still emerging. Enhancing
joint sensing with AI techniques, considering UAV resource limits,
and developing power-efficient routing protocols for varying channel
conditions are also key future directions. Implementing state-of-the-art
communication technologies such as 6G, cloud and fog computing could
lead to faster communication, allow for higher levels of autonomy
and promote mission success. Moreover, improving hardware platforms
(e.g., FPGA-based (OcPoC-Zynq Mini), ARM-based (Pixhawk), GPU-based
platforms (Nvidia Jetson)) will significantly enhance embedded processing
units and flight controller capabilities. Developing high-speed onboard
processing units with substantial memory capacity will be vital for
handling the complex computations required in autonomous UAV operations.

\paragraph*{Planning and Perception}Perception and UAV path planning
remain complex, multi-faceted challenges. Perception is enabled using
multi-sensor fusion. However, multi-sensor fusion faces three main
challenges, namely, sensor failure, synchronizing heterogeneous signals
and updating state localization using sensor measurements obtained
at varying frequencies. Therefore, future research should focus on
fault management and redundancy architectures to address signal denial
or sensor failures. Although the sensor fusion algorithms mentioned
above demonstrate high accuracy in object detection and classification,
their performance and reliability under extreme weather conditions
and varying lighting require further evaluation. Additionally, future
efforts should develop perception techniques capable of synchronizing
heterogeneous signals and handling varying sensor measurement frequencies.
Most reported environment modeling techniques are applied primarily
to simpler types of environments, mostly known and static with centralized
control. Centralized models raise concerns about the scalability of
planning approaches, even in the absence of a centralized topology.
This can lead to each UAV waiting for its peers to finish their planning,
creating a potential computational bottleneck in large UAV systems.
Game theory will gain more popularity for routing, optimizing energy
consumption, network coverage, resource management, coordinated control
in UAV networks, and UAV cluster path planning.

\paragraph*{Onboard UAV Limited Computational Resources}Perception,
estimation, path planning, control, communication, and classification
algorithms that rely on advanced machine learning techniques and computationally
intensive training can present significant challenges due to the limited
computational resources available on the UAV's onboard system. This
issue of lengthy computational training can be addressed by outsourcing
the training phase to the ground, which simplifies the algorithm to
a basic mapping function%
. However, this approach is limited because it prevents any new learning
from being introduced online. As a result, the development of computationally
efficient machine learning algorithms that can be implemented onboard
a drone as well as outsource training becomes a crucial research direction
for UAVs. In case of outsourcing, ensuring privacy and security is
essential to prevent breaches, adversarial attacks, theft, or unauthorized
access.

\paragraph*{Data Securing}As UAV networks expand, the volume of real-time
data streaming exchanged between UAVs and ground stations also increases.
Hence, large volumes of data must be securely aggregated to protect
against malicious threats. Robust encryption methods are essential
to ensure confidentiality during data transmission between GCS and
UAVs. Implementing secure and efficient data aggregation techniques
can lower communication costs and energy use while safeguarding data
privacy. Ongoing research and innovation are crucial to strengthen
UAV systems' resilience against potential attacks. While this is an
area of ongoing development, it continues to be a challenging open
problem. Blockchain technology, as a decentralized solution, is emerging
as a transformative tool expected to revolutionize privacy protection
for UAV systems by providing enhanced security and adaptability. Aerial
blockchain can safeguard UAV communication privacy and ensure the
integrity of data collected by UAVs. Additionally, integrating blockchain
into UAV softwarization can enable dynamic, flexible, and real-time
communication services within UAV networks.

\paragraph*{Regulation}Current UAVs feature diverse hardware and
software, but their success relies heavily on local regulations and
faces challenges related to global collaboration and standardization.
Initiatives like Joint Authorities for Rule-making on Unmanned Systems
(JARUS) and Single European Sky ATM Research (SESAR) play crucial
roles in shaping UAV development globally. Autonomy is a major innovation,
offering benefits such as pilot-free operations, greater efficiency,
and lower costs. Technologies like UAV traffic management and Beyond
Visual Line of Sight (BVLOS) piloting are gaining traction, potentially
transforming the industry. SESAR, a key European regulatory body,
aims to enhance airspace safety, efficiency, and environmental impact
through advanced technologies, positioning itself as a leader in autonomous
drone development. There is a growing need for flexible, risk-based
UAV regulations that consider factors like area, purpose, and visibility.
In Canada, UAVs between 250g and 25 kg must be registered, and pilots
certified. Operations with UAVs over 25 kg in Canada require a Special
Flight Operations Certificate (SFOC), adding administrative burdens
that can deter hobbyists and commercial users. In contrast, EASA adopts
a risk-based approach with fewer regulations for low-risk operations.
Balancing the needs of stakeholders, regulators, manufacturers, and
users, is crucial. Future regulations should account for UAV weight,
payload, and operational parameters beyond just weight, reflecting
their diverse applications.

\paragraph*{More Dataset}Traditional machine learning methods rely
on a centralized data source for model training, whereas Federated
Learning (FL) and Federated Deep Learning (FDL) involves multiple
entities collaboratively training a model. FL and FDL are areas of
active research in the context of UAVs due to its potential to enhance
data privacy and support network scalability as the number of UAVs
increases. However, a significant challenge hindering the immediate
deployment of FL models on UAVs is the limited availability of relevant
datasets for training, such as network traffic and malware datasets%
. As a result, future research is likely to focus on generating more
datasets to facilitate effective model training%
{} which are helpful for various UAV avionics systems, including electronic
warfare countermeasures (e.g., explainable cyberattacks and spoofing
detection). For instance, cyberattacks may not be properly classified
due to insufficient training datasets. Moreover, while trained models
often detect cyberattacks, their classifications are not always explainable.
Understanding and explaining these attacks is crucial for designing
robust countermeasure techniques.

\section{Conclusion\label{sec:Conclusion}}

The advancement of avionics systems for Unmanned Aerial Vehicles (UAVs)
has a direct impact on enhancing UAV decision making and autonomous
mobility capabilities. This paper provides a detailed overview of
the most crucial components of UAV avionics. The role of UAV communication
systems, antennas characteristics and selection, and techniques for
sharing UAV's location have been reviewed. Identification systems
and their role in responding to interrogating signals have been discussed.
Power sources taxonomy and selection based on UAV size have been covered.
Perception sensors, sensor fusion, navigation techniques, and the
importance of filter design for UAV autonomy have been studied. Common
path planning techniques and their interrelation with the collision
avoidance methods, and state-of-the-art tracking control techniques
have been presented. Electronic warfare threats and methods including
destructive and non-destructive cyberattacks, transponder attacks,
jamming threats, countermeasures and defensive aids approaches have
been explored. Moreover, the role that different types of databuses
play in enabling efficient and fast real-time data transfer has been
discussed. This overview has been concluded by outlining how safety
is incorporated into the UAV design, safety occurrence trends, and
the associated national regulations along with the certification process.

\section*{Acknowledgments}

This work was supported in part by the National Sciences and Engineering Research Council of Canada (NSERC), under the grants RGPIN-2022-04937. The author would like to thank \textbf{Maria Shaposhnikova} for proofreading the article.

\balance
\bibliographystyle{IEEEtran}
\bibliography{bib_Avionics_UAV}
\end{document}